\documentclass[%
 aip,cha,
 amsmath,amssymb,
reprint,%
floatfix,%
]{revtex4-1}

\usepackage{graphicx}
\usepackage[font={large,sf,bf,sc}]{subfig} 
\usepackage{dcolumn}
\usepackage{bm}
\usepackage[mathlines]{lineno}

\usepackage[utf8]{inputenc}
\usepackage[T1]{fontenc}
\usepackage{mathptmx}
\usepackage{etoolbox}
\usepackage{xcolor} 
\usepackage[section]{placeins}
\usepackage[font=small,labelfont=bf,
justification = justified,format=plain]{caption}

\usepackage{float}

\newcommand{\red}[1]{\textcolor{red}{#1}}
\newcommand{\blue}[1]{\textcolor{blue}{#1}}
\newcommand{\avg}[1]{\left\langle#1\right\rangle} 
\newcommand{\tspk}{t_{\rm spk}} 
\newcommand{\Iext}{I^{\rm ext}} 
\newcommand{\Isyn}{I^{\rm syn}} 
\newcommand{\Qfast}{Q_{\rm fast}} 

\makeatletter
\def\@email#1#2{%
 \endgroup
 \patchcmd{\titleblock@produce}
  {\frontmatter@RRAPformat}
  {\frontmatter@RRAPformat{\produce@RRAP{*#1\href{mailto:#2}{#2}}}\frontmatter@RRAPformat}
  {}{}
}%
\makeatother

\newlength\savedwidth

\newcommand\thickhline{\noalign{\global\savedwidth\arrayrulewidth\global\arrayrulewidth 2pt}%
\hline
\noalign{\global\arrayrulewidth\savedwidth}}

\newcommand{\adicionado}[1]{#1}

\begin{document}

\preprint{AIP/123-QED}

\title{Optimal input reverberation and homeostatic self-organization towards the edge of synchronization}
\author{Sue L. Rhamidda}
\affiliation{ 
Departamento de Física, FFCLRP, Universidade de São Paulo, Ribeirão Preto, SP, 14040-901, Brazil}
\author{Mauricio Girardi-Schappo}
\affiliation{Departamento de Física, Universidade Federal de Santa Catarina, Florianópolis, SC, 88040-900, Brazil}
\author{Osame Kinouchi}%
 \email{suelam@usp.br, girardi.s@gmail.com, okinouchi@gmail.com}
\affiliation{ 
Departamento de Física, FFCLRP, Universidade de São Paulo, Ribeirão Preto, SP, 14040-901, Brazil}


\date{\today}

\begin{abstract}
Transient or partial synchronization can be used to do computations, although a fully synchronized network is sometimes related to the onset of epileptic seizures.
Here, we propose a homeostatic mechanism that is capable of maintaining a neuronal network at the edge of a synchronization transition,
thereby avoiding the harmful consequences of a fully synchronized network.
We model neurons by maps since they are dynamically richer than integrate-and-fire models and more computationally efficient than
conductance-based approaches.
We first describe the synchronization phase transition of a dense network of neurons with different tonic spiking frequencies coupled by gap junctions.
We show that at the transition critical point, inputs optimally reverberate through the network activity
through transient synchronization.
Then, we introduce a local homeostatic dynamic in the synaptic coupling and
show that it produces a robust self-organization toward the edge
of this phase transition.
We discuss the potential biological consequences of this self-organization process, such as its relation to the Brain Criticality hypothesis,
its input processing capacity, and how its malfunction could lead to pathological synchronization \adicionado{and the onset of seizure-like activity}.
\end{abstract}

\maketitle

\section{Introduction}

\textbf{Synchronization \adicionado{can be seen as} the coincidence of spike times in a given population of interacting neurons~\cite{Brette2012}.
A consequence of synchronous spiking is the emergence of collective oscillations in the network.
Transient synchronicity can be used for computation~\cite{Izhikevich2006,Brette2012,Palmigiano2017}. However,
a fully synchronous network -- one in which the voltage oscillates with high amplitude and frequency -- can be regarded as a seizure model~\cite{Lehnertz2009Epi,Rich2020Epi}.
This pathological state could be the result of an intricate play of time scales~\cite{jirsaEpileptor2014},
and disruption of neuromodulation could generate some seizures by excessive excitation~\cite{LopesdaSilva2003}.
In this paper, we study a system with optimal input reverberation at the critical synchronization point.
When subjected to a homeostatic mechanism, this dynamical framework could be used to understand some routes to the generation of seizure\adicionado{-like activity}.}

The Brain Criticality hypothesis states
that the critical point is central to the processing of information in the healthy brain~\cite{Beggs2003,Kinouchi2006,Beggs2008,Shew2013}.
Several studies show the presence of a critical state, with power-law-distributed neuronal avalanches and other
scaling properties (for reviews, see~\cite{Cocchi2017,Girardi2021,plenz2021self,Obyrne2022}).
The originally observed critical point was conjectured to separate an inactive state from a synchronous epileptic-like state~\cite{Beggs2003},
although most of the theoretical works that followed relied on absorbing~\cite{Carvalho2021} or Ising universality classes~\cite{PonceDeco2018zfish}
(a continuous phase transition between an inactive/disordered and an active/ordered state, where there is no synchronization).
In such models, homeostatic mechanisms in synaptic coupling~\cite{Levina2007},
in firing gain~\cite{Kinouchi2019}, or in multiple variables~\cite{Girardi2020,Girardi-Schappo2021,Menesse2021}
lead to hovering around the critical point~\cite{Kinouchi2020,Chialvo2020,Buendia2020}.

The connection between power-law avalanches and a synchronization phase transition has also been
investigated~\cite{Poil2012,DiSanto2018,DallaPorta2019,Buendia2021}.
The authors used simplified models of phase oscillators~\cite{Buendia2021},
integrate-and-fire (IF) neurons~\cite{Poil2008,DallaPorta2019}, or abstract population equations~\cite{DiSanto2018}.
However, the question remains whether the homeostatic dynamics could lead to the edge of synchronization,
similarly to what happens with absorbing/Ising phase transitions.

\begin{figure}[bp!]
    \centering    
    \includegraphics[width=0.4\textwidth]{{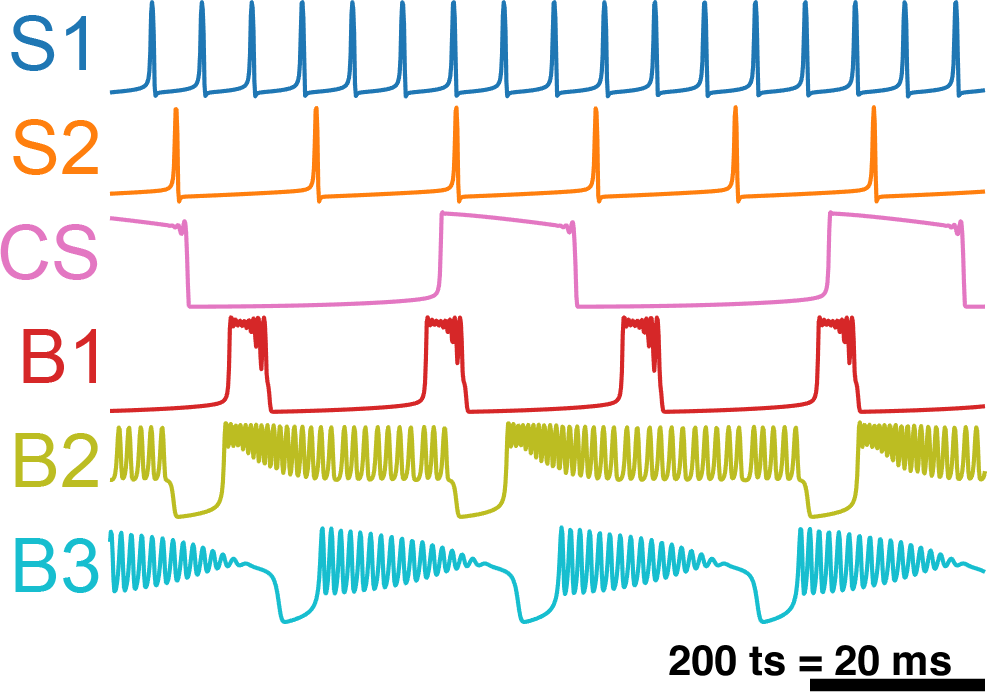}}
    \caption{\label{fig:spike}\adicionado{
    \textbf{Example attractors of the KTH neuron.} Parameters: see Table~S1 in the Supplementary Information.
    10 time steps=1ms. The spikes are a result of the underlying interplay of fast (given by the $T$ and $K$ parameters; Eq.~\eqref{eq:kthmodeldef}) 
    and slow time scales ($\delta_i$ and $u$ parameters), analogously to what happens in conductance-based neuronal models.
    \textbf{S1, S2}: tonic spiking;
    \textbf{CS}: cardiac spiking;
    \textbf{B1, B2, B3}: dynamically distinct bursts.}}
\end{figure}

\begin{figure*}[!tp]
    \centering    
    \includegraphics[width=\textwidth]{{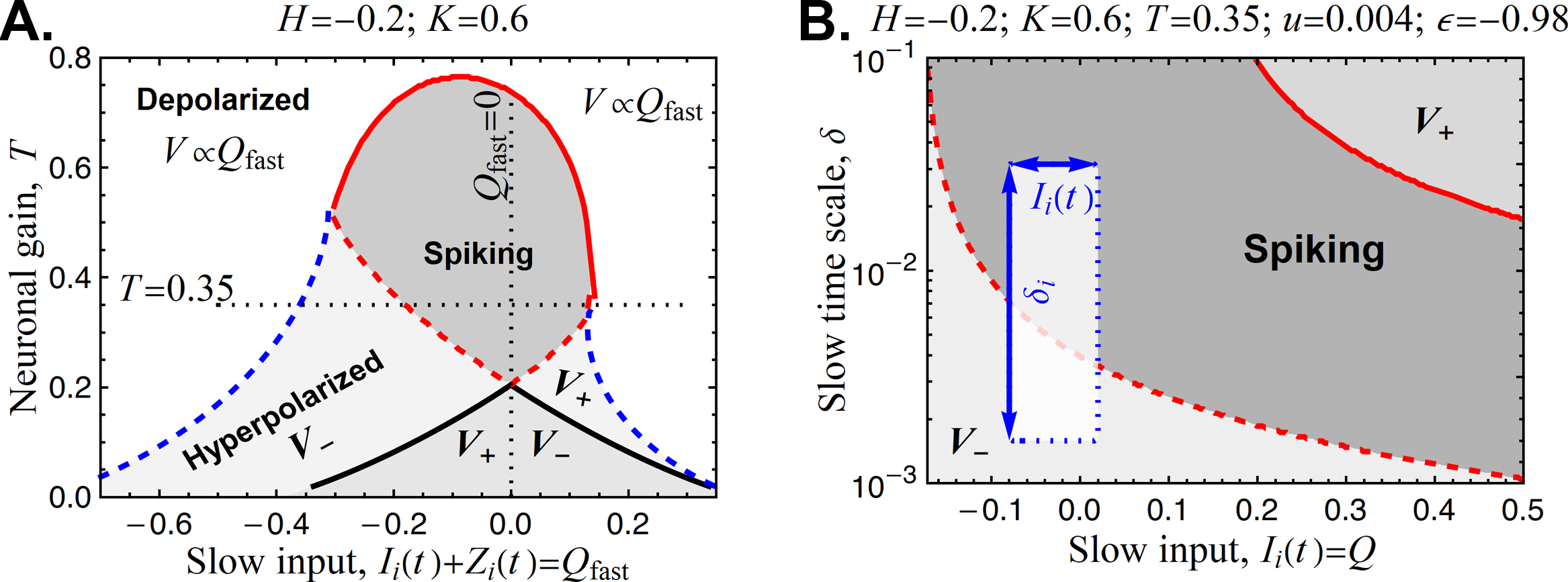}}
    \caption{\label{fig:phasediag}\adicionado{
    \textbf{Phase diagrams of an isolated neuron.} $V$ is
    the fixed point of the map, and $V_+$ and $V_-$ are the hyperpolarized membrane potentials
    related to an upper ($V>0$) and a lower ($V<0$) stable fixed points.
    The \textit{Spiking} phase has only stable limit cycles with periods that depend
    on $Q$, $T$, and $\delta$.
    For both panels: (\textcolor{black}{\textbf{-----}}) stable fold bifurcations of the upper ($V_+$) and lower ($V_-$) fixed points;
                     (\blue{\textbf{- - -}}) unstable fixed point fold bifurcation;
                     (\red{\textbf{- - -}}) homoclinic bifurcation of limit cycle (infinite period bifurcation; stable fold point);
                     (\red{\textbf{-----}}) supercritical Neimark-Sacker bifurcation.
    \textbf{A.} Fast subsystem phase diagram obtained by replacing both the slow current $Z_i(t)$ and the input $I_i(t)$ with a constant
    field $\Qfast=Z_i(t)+I_i(t)$. For a fixed $T$, the $Z_i(t)+I_i(t)$ currents cause the state of the neuron to oscillate left
    and right in this graph.
    \textbf{B.} Phase diagram of the full neuron replacing only the input $I_i(t)$ with a constant field $Q$. The highlighted rectangle (height=$2\Delta$; width=$I_i(t)$ amplitude)
    illustrates the state of a network of $N$ neurons with uniformly distributed $\delta_i$. The effect of the synaptic current
    is making a given neuron (with fixed $\delta_i$) oscillate left and right within this rectangle. The amplitude of $I_i(t)$ has been exaggerated for visualization,
    since it is of the order of $10^{-4}$ (Fig.~\ref{fig:syncraster}).
    The lines are the stability limits of the fixed point
determined by the linear stability analysis described
elsewhere~\cite{Kinouchi1996,Girardi2017}.
    }}
\end{figure*}

Here, we introduce a network made up of map neurons under the influence of a simple rule of synaptic plasticity
that is capable of pushing the system toward the edge of a synchronization phase transition.
In particular, our map-based neurons offer a trade-off between analytical tractability,
computational efficiency, and a rich repertoire of dynamical behaviors~\cite{Courbage2010,Ibarz2011,Girardi2013}.
Unlike IF models, spikes appear naturally \adicionado{with a stereotyped waveform} on maps as a consequence of the interplay between its time scales,
yielding a dynamical picture that is generally similar to that of conductance-based models~\cite{Girardi2017}.
Contrary to previous similar maps~\cite{Kinouchi1996,Kuva2001,Copelli2004,Girardi2013b,Girardi2017},
ours provides better control of the tonic spiking frequency -- an important feature for
the present study.

The synchronization phase transition we found has a power-law decay in the amplitude of the activity following stimuli due
to critical slowing down.
This yields optimal reverberation of the inputs.
Some bifurcations with this feature are interpreted as
second-order (critical) phase transitions in generalized Ising models~\cite{Yokoi1985,Tragtenberg1995}.

We explore the role of the homeostasis parameters -- henceforth called
``hyperparameters'' -- and show that the convergence to the
critical point is robust. This is because the system still reaches
the vicinity of the critical point after either perturbing the synapses or changing the hyperparameters within a large range.
The change in these homeostatic parameters implies only gross tuning for living systems.
Our model provides a simple \adicionado{mechanistic} explanation for \adicionado{two routes to} the genesis of seizure\adicionado{-like activity}:
one is the disruption of the homeostatic dynamics and the second is the momentary synchrony due to external stimuli.
Thus, such dynamics around the synchronization phase
transition could have consequences for the maintenance of
the healthy brain state (such as avoiding seizures~\cite{LopesdaSilva2003}) and its processing of
inputs~\cite{Brette2012}.

\section{Results}

\begin{figure*}[!tp]
    \centering     
    \includegraphics[width=\textwidth]{{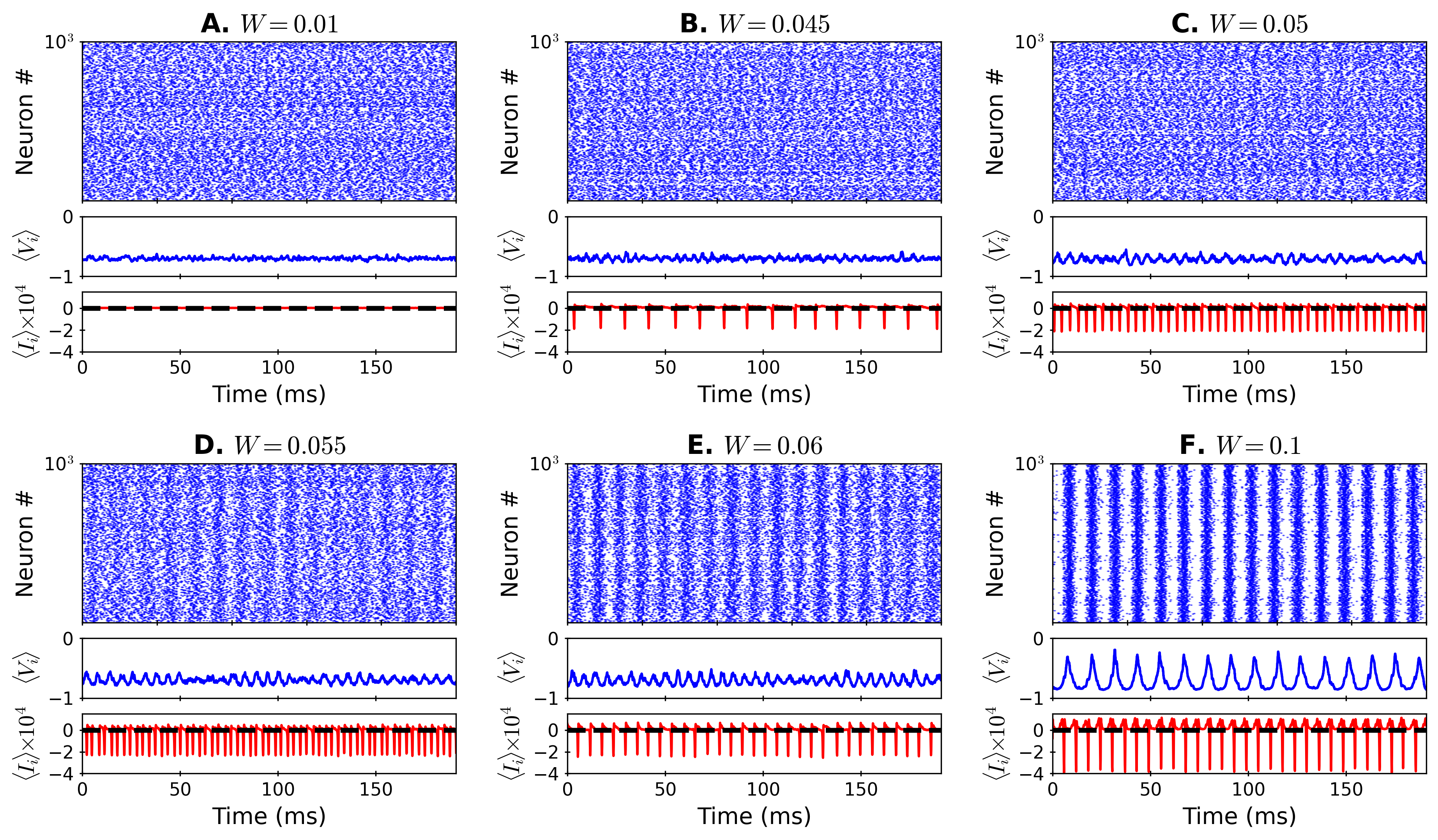}}
    \caption{\label{fig:syncraster}
      \textbf{Network activity as a function of $W$.}
      All panels: raster plot showing spikes $S_i[t]$ (top); mean network potential (middle);
      and mean input $\avg{I_i[t]}\times10^4$ (bottom). (\textbf{- - -}) mean over time, $\avg{\avg{I_i[t]}}_t$. Notice that the mean input is of the order $10^{-4}$ regardless
of the coupling intensity. Parameters:
      $N=1000$, $K=0.6$, $T=0.35$, $H=-0.2$, $\delta=0.006$, $\Delta=0.003$, $u=0.004$, $\epsilon=-0.98$.}
\end{figure*}

First, we present \adicionado{a new} neuron map model. Then, we examine its collective synchronization phase transition and show that there is optimal reverberation
of inputs at the critical point. Finally, we propose a synaptic mechanism that
\adicionado{avoids synchronization,} leading the network to lurk around the critical point vicinity.
\adicionado{We then probe for its robustness and discuss how its disruption may trigger the onset of seizure-like activity via two different routes}.

\subsection{The map-based neuron model}
\label{sec:model}

The membrane potential $V_i(t)$ as function of time of an isolated neuron $i$ is obtained by iterating the following equations:
\begin{equation}
\label{eq:kthmodeldef}
\begin{array}{lll}
   V_{i}[t+1] & = &  \tanh\!\!\left( \dfrac{V_{i}[t]-K Y_i[t] + Z_{i} [t] + I_i[t] }{T}\right) \:, \\
   Y_{i}[t+1] & = &  \tanh\!\!\left( \dfrac{V_{i}[t] + H}{T} \right)  \:, \\
   Z_{i}[t+1] & = &  Z_{i}[t] - \delta_i Z_i[t] - u \left(V_{i}[t] - \epsilon\right) \:,
\end{array}
\end{equation}
where the $t$ index denotes discrete time steps;
$K$, $T$, $H$ are parameters that control
the fast dynamics;
$\delta_i$, $u$, $\epsilon$ control
the slow variable $Z_i[t]$ ($\delta_i \ll 1$).
The synaptic and external inputs enter in the
current $I_i[t]=\Isyn_i[t]+\Iext_i[t]$.
A typical spike takes roughly 10 time steps,
which then corresponds to 1~ms. We consider 
$\Iext_i[t]=0$ unless otherwise specified.

\adicionado{This map was proposed in~\cite{kinouchi2001map} and called the KTH model, although its dynamic
has not been described}.
The $Y$ variable differs from previous models~\cite{Girardi2013},
making the fast subsystem more diverse.
The membrane potential $V_i(t)$ is always bounded within $(-1,+1)$.
This neuron can display fast and slow spiking and bursting, and
plateau spikes/bursts, with variable interspike and interburst
intervals. It is similar to the Hindmarsh-Rose equations~\cite{Hindmarsh1984}.
Some exemplars \adicionado{of the spiking behavior of a KTH neuron} are shown in Fig.~\ref{fig:spike}.

\adicionado{The parameters $T$ and $K$ modulate a fast self-coupling, and thus can be interpreted as a fast Sodium dynamics,
since they are responsible for generating the spike depolarization.
The slow $Z_i$ current, on the other hand, is responsible for modulating the fast dynamics, and can be interpreted
as a slow Potassium feedback with normalized inverse time constants $\delta_i$ and $u$ and reversal potential $\epsilon$.}

\adicionado{We perform a slow-fast analysis of the map to unveil the bifurcations that lead to spiking,
and consequently generate the synchronization phase transition to be studied ahead.
It consists of substituting the slow variable by a fixed parameter, and tracing the bifurcation
diagrams as a function of this new parameter -- a procedure that is also know as \textit{adiabatic approximation}~\cite{Copelli2004}.
The bifurcations were identified via standard linear stability analysis (we followed the exact procedure described elsewhere~\cite{Girardi2017}).}

\adicionado{To get a glimpse on the fast dynamics, we define the constant $\Qfast\equiv Z_i+I_i$ replacing both the $Z_i$ current
and the external input, and trace the bifurcation diagram in Fig.~\ref{fig:phasediag}A.
The spiking regime appears through an infinite-period (homoclinic) bifurcation or a supercritical Neimark-Sacker bifurcation,
depending on the value of $\Qfast$.
We also identified stable fold bifurcations that give rise to the coexistence of hyperpolarized states $V_+>0$ and $V_-<0$
corresponding to stable fixed points.
For a fixed $T$, the $Z_i(t)+I_i(t)$ currents cause the state of the neuron to oscillate left and right crossing
the bifurcation lines and generating bursting (if $T\gtrsim0.2$; \textit{i.e.}, around the \textit{bubble}) or cardiac spikes (via self-driven hysteretic cycles if $T\lesssim0.2$;
below the \textit{bubble}).}

\adicionado{We now replace only the external input with a constant parameter $Q\equiv I_i$ in order
to understand the how the neurons will interact with one another when coupled in the network (Fig.~\ref{fig:phasediag}B).
In this case, the lower and upper fixed points, $V_-$ and $V_+$ respectively, are separated by an intermediary spiking phase.
For a fixed $\delta$, the negatively hyperpolarized membrane, $V_-$, undergoes an infinite-period bifurcation as $Q$ increases.
In other words, if the neuron is sufficiently close to this bifurcation curve, then increasing external inputs
are capable of generating both slow spiking (near the bifurcation) and fast spiking (away from the infinite-period bifurcation,
but before crossing to the $V_+$ region). 
This is because the spiking frequency \adicionado{is known to have} a power law divergence \adicionado{at the homoclinic bifurcation line}~\cite{strogatzBook,Girardi2017}.
The proximity to the homoclinic bifurcation is behind the emergence of synchronization
that we will study in the next sections.}

\subsection{The synchronization transition}

\adicionado{All the results that follow were obtained for networks of coupled KTH maps.}
We build an all-to-all network of $N$ neurons with diffusive coupling through gap junctions~\cite{RothRossum2010}.
The synaptic input on neuron $i=1,\cdots,N$ is:
\begin{equation}
\label{eq:gapjunction}
    \Isyn_i[t] = \dfrac{W}{N} \sum_{j\neq i} (V_j[t]-V_i[t])\:,
\end{equation}
$W$ is the synaptic weight (assumed homogeneous for simplicity),
and the sum runs over all the presynaptic neighbors $j$.
\adicionado{We could have chosen to use heterogeneous coupling weights $W_{ij}$ instead,
as long as the weights obey a self-averaging distribution with mean $W=\avg{W_{ij}}$ taken over the off-diagonal pairs.
In this case, the control parameter is still the mean synaptic weight $W$.}
Eq.~\eqref{eq:gapjunction} is a linearized Kuramoto interaction~\cite{Kuramoto1984}.


\adicionado{All parameters (except for $\delta_i$) are homogeneous in the network.
All neurons are set in a tonic spiking mode \adicionado{(exemplified in Fig.~\ref{fig:spike} with label S2)}.
The parameter $\delta_i$ controls the recovery time scale of the slow variable $Z_i[t]$. Smaller
$\delta_i$'s imply in slower recovery times. Thus, this parameter is intrinsically related to the natural spiking frequency of
each neuron. We force heterogeneity in the network by making each neuron $i$ to have a unique $\delta_i$
uniformly chosen in the range $\left[\delta-\Delta;\delta+\Delta\right]$, $\avg{\delta_i}=\delta$.
The resulting population is mixed in the sense that some neurons (with smaller $\delta_i$) require more input than others (with larger $\delta_i$)
to cross the bifurcation line -- see the highlighted rectangle in Fig.~\ref{fig:phasediag}B
representing a population of neurons.
Thus, for $W=0$, neurons are independent, but still fire regularly, yielding a non-vanishing
global firing rate $\rho$ (spikes per neuron in the stationary state).
This suggests that a synchronization phase transition could occur for increasing $W>0$
when a set of $N$ neurons interact near the edge of this bifurcation.}

A perfectly synchronized network has
site-averaged membrane potential $\avg{V_i[t]}\equiv V[t]$
equal for all neurons, $V[t]=\avg{V_i[t]}=\avg{V_j[t]}$, making $\avg{I_i[t]}=0$.
However, the amplitude of the total synaptic input is proportional to $W$,
which then determines whether a given neuron $i$ enters
or leaves the spiking region of the phase diagrams in Fig.~\ref{fig:phasediag}.
Thus, $W$ is the control parameter for the synchronization transition.

We define an arbitrary threshold $\lambda$ in order to obtain a binary spike variable $S_i[t]$,
such that $S_i[t=\tspk]=1$ if $V[\tspk]\geq\lambda$, then a spike occurred at $t=\tspk$; $S_i[t]=0$ otherwise.
The \adicionado{instantaneous} firing rate of the network is defined as the fraction of spikes
per unit of time, $\rho[t]=\avg{S_i[t]}=(1/N)\sum_iS_i[t]$.

Fig.~\ref{fig:syncraster} shows the activity as $W$ is increased,
suggesting that there is a synchronization critical point $W_c$.
\adicionado{Note that the time-averaged input current $\avg{\avg{I_i[t]}}_t$ remains
close to zero (of order $10^{-4}$), although its amplitude increases allowing for neurons to spike
together, as predicted}. Collective oscillations with large amplitude (of order 1)
emerge on the average potential $V[t]$ and consequently on the firing rate $\rho[t]$.
The $\avg{\cdots}_t$ represents the temporal average over a long time interval
in the stationary state of the network.

Moreover, the raster plots show that neurons do not spike regularly within the synchronized state,
making this a synchronous-irregular phase~\cite{Brunel2000,Girardi2020} -- \adicionado{see the interspike interval
distribution for a single neuron in the network in Supplementary Fig.~S1.
Now, we characterize this phase transition by defining an order parameter,
and then show that it optimizes the reverberation of inputs in the network.}

\subsection{Order parameter}

\begin{figure}[!tp]
    \centering
    \includegraphics[width=0.4\textwidth]{{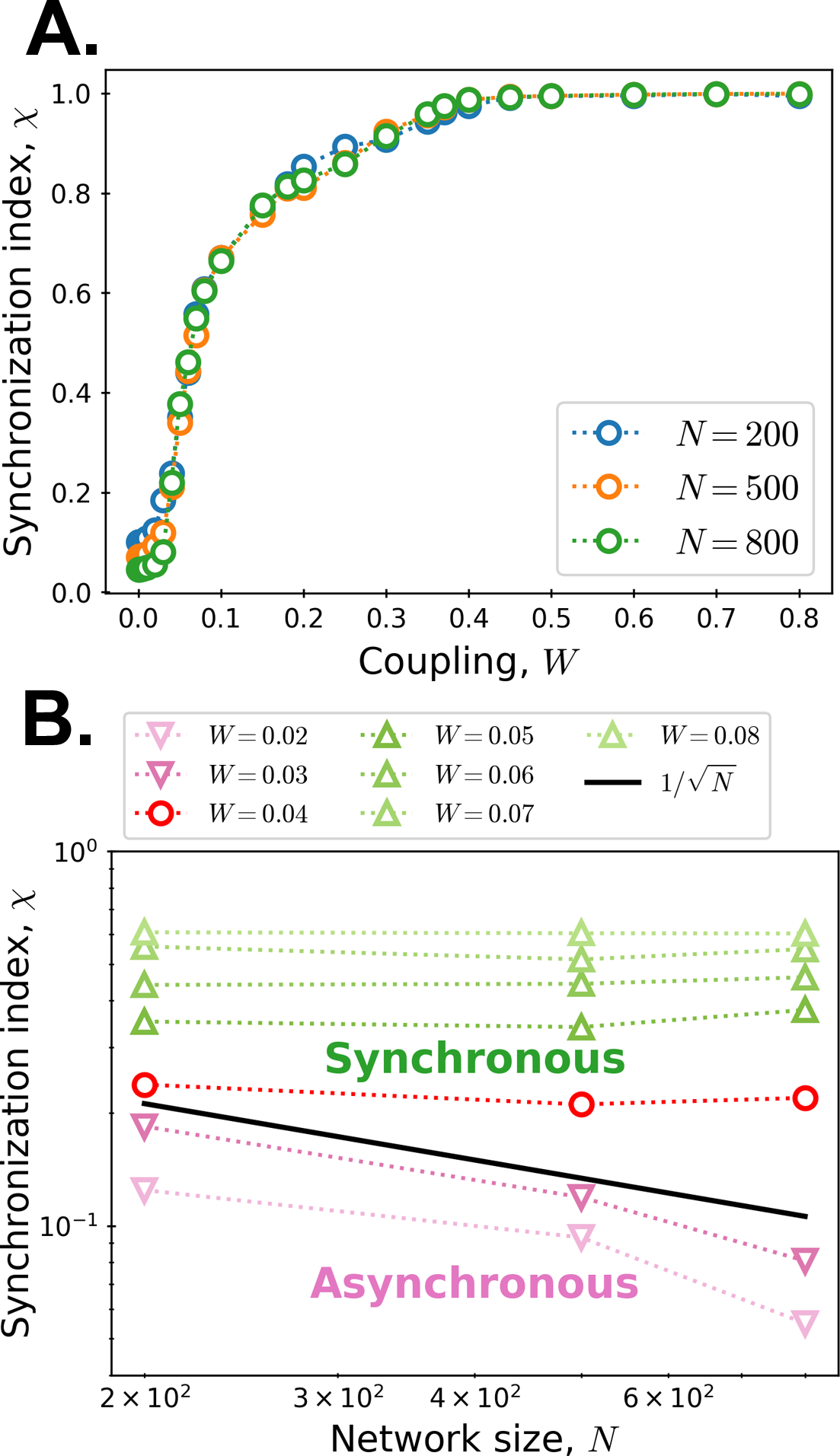}}
    \caption{\label{fig:csi}
    \textbf{Synchronization index as a function of coupling and network size.}
    \textbf{(A)} $\chi$ rises as the network becomes synchronized with increasing $W$.
    \textbf{(B)} For $W\lesssim0.04$, $\chi$ decays with the scaling $1/\sqrt{N}$ or faster,
    showing that the phase transition happens around $W_c\approx0.04$, as expected.
    Due to the finite-size of the network, we consider the critical point to be in the range $0.04\leq W_c\leq0.05$.
    Parameters: $K=0.6$, $T=0.35$, $H=-0.2$, $\delta=0.006$, $\Delta=0.003$, $u=0.004$ and $\epsilon=-0.98$.}
\end{figure}

\begin{figure*}[!tp]
      \centering
      \includegraphics[width=\textwidth]{{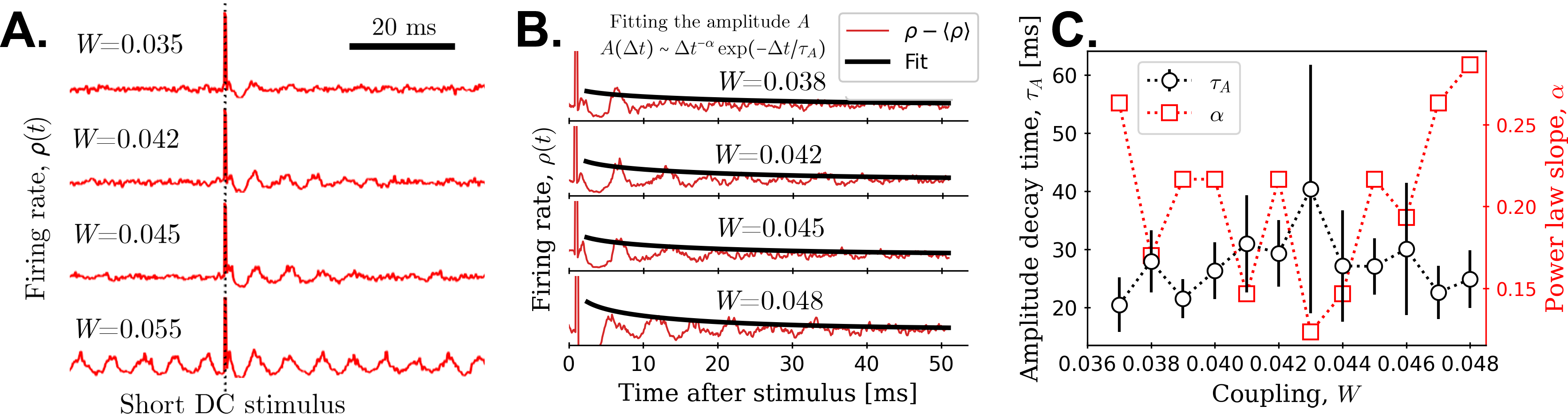}}
      \caption{\label{fig:reverbfit}
    \textbf{Reverberating response to external stimuli near synchronization}.
    \textbf{A.} Oscillations are damped for $W\lesssim0.050$. \adicionado{For $W>0.050$, the input only quickly perturbs the oscillation mode of the network.}
    \textbf{B.} We estimated the characteristic time of the system by fitting the damping of the amplitude of the oscillations (see Supplementary Information for details).
    \textbf{C.} The characteristic time $\tau_A$ of the best fit has a maximum near the critical point due to critical slowing down.
    The shown $\alpha$ minimizes the fitting error of $\tau_A$ for each $W$; it does not have error bars because it is a fixed parameter during the fitting procedure.
    $\tau_A$ is measured on the left-hand scale, and $\alpha$ is measured on the right-hand scale.}
\end{figure*}

\adicionado{
We employ the synchronization index to identify the phase transition~\cite{Golomb1994,Golomb2007},
since there is no vanishing activity -- \textit{i.e.}, absorbing state -- near the transition point.
The site-averaged network membrane potential is
\begin{equation}
\label{eq:meanV}
    \avg{V_i[t]}\equiv V[t] = \frac{1}{N} \sum_i V_i[t]\:.
\end{equation}
Making $\avg{\cdots}_t$ the average over time in the stationary state, the temporal variance of the network voltage is:
\begin{equation}
\sigma_V^2 = \avg{V[t]^2}_t-\avg{V[t]}_t^2\:,
\end{equation}
whereas for a given neuron $i$ in the network, we have:
\begin{equation}
\sigma^2_{V_i} =  \avg{V_i[t]^2}_t-\avg{V_i[t]}_t^2\:,
\end{equation}
leading to the synchronization index
\begin{equation}
\chi^2(N) =  \dfrac{\sigma^2_V}{\avg{\sigma^2_{V_i}}} \:.
\end{equation}
The global variance is normalized by the
average of the individual variances of the neurons.
Thus, when $N\to\infty$, the index goes from
 $\chi\sim1/\sqrt{N}$ in
 the asynchronous state to $\chi=1$ for complete synchronization~\cite{Golomb2007}.
 This is because adding $N$ temporally uncorrelated (or weakly correlated) signals
results in a population average whose fluctuations have an
amplitude of order $1/\sqrt{N}$.
}

There is a phase transition near $W=W_c\approx0.045$ where
incipient oscillations are forming.
$\chi$ continuously grows with increasing coupling $W$, and the scaling $1/\sqrt{N}$ occurs for $W\lesssim0.04$ (Fig.~\ref{fig:csi}).
Finite size effects prevent $\chi=0$ even at $W\leq0.04$, so in our simulations we consider that the critical point
lies in the interval $0.04\leq W_c\leq0.05$, and estimate it to be $W_c=0.045$.

\subsection{Optimal input reverberation at criticality}

We further characterize the phase transition by its response to external inputs near $W_c$.
In the following, we show evidence that $W_c$ optimizes the reverberation of the input
through critical slowing down.
\adicionado{We stimulate the network with a short square DC pulse input only} after reaching the stationary state.
The input is applied to 30\% of the network \adicionado{and can be written as}
\begin{equation}
\label{eq:inputpulse}
    \Iext_i[t]=I_0\Theta[t\adicionado{-t_0}]\adicionado{\equiv I_0\Theta[\Delta t]}\:,
\end{equation}
where $I_0=0.3$ is the amplitude and $\Theta[\Delta t]$ is a normalized square pulse (lasting $\sim1$~ms),
\adicionado{and $\Delta t=t-t_0$ is the delay after the stimulus is injected at $t_0$}.
This forces a part of the network to spike together,
momentarily synchronizing \adicionado{the few randomly selected neurons}.

Fig.~\ref{fig:reverbfit}A shows the network activity $\rho[t]$ responding with damped oscillations \adicionado{both at the critical point $W_c$ and
in the subcritical state, $W<W_c$}. Conversely, the synchronous state is only \adicionado{instantaneously} perturbed,
\adicionado{quickly} returning to its previous persistent oscillation \adicionado{mode}.

\adicionado{We define the reverberation of the input as the transient synchronization that appears
after the stimulus stops~\cite{Palmigiano2017}. The duration of the transient can be defined as the characteristic time for the damping
of the amplitude $A$. Thus,}
we fitted the envelope of the oscillatory decay (Fig.~\ref{fig:reverbfit}B \adicionado{-- for full details, see the
Supplementary Information}) using the function
\begin{equation}
    \label{eq:rhoamp}
    A[\Delta t]\sim\Delta t^{-\alpha}\exp\!\left(\Delta t/\tau_A\right)\:,
\end{equation}
where $\tau_A$ is the damping characteristic time fitted using nonlinear least squares; $\Delta t$ is the
delay after the stimulus (not to be confused with the $\Delta$ parameter); $\alpha$ is a critical exponent.
We fit Eq.~\eqref{eq:rhoamp} for fixed $\alpha$ to obtain one $\tau_A$ for each $\alpha$. We do this for many $\alpha$
and select the $\tau_A$ that minimizes the fitting error over $\alpha$. See the Supplementary Figs.~S2,S3 for details.
\adicionado{We included} the power-law term \adicionado{($\Delta t^{-\alpha}$)} \adicionado{inspired by two things. First,}
the \adicionado{typical amplitude decay in a} supercritical Neimark-Sacker bifurcation
as the critical point $W_c$ is approached \adicionado{has the form} $A\sim|W-W_c|^{-1/2}$.
\adicionado{Secondly, we showed that the isolated KTH map presents supercritical Neimark-Sacker bifurcations in the slow-fast analysis.
Eq.~\eqref{eq:rhoamp} also yielded less error when compared to a purely exponential fitting model,
so we kept the power law.}

\begin{figure*}[!htp]
    \centering
    \includegraphics[width=\textwidth]{{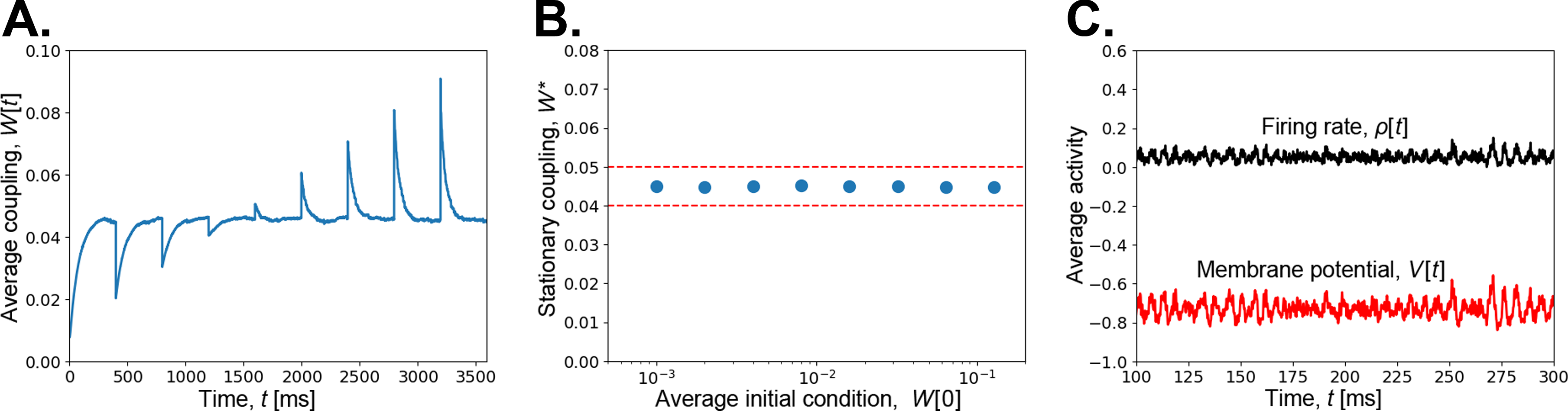}}
    \caption{\label{fig:Wt}
    \textbf{Synchronization-avoiding homeostatic plasticity robustness.}
    \textbf{A.} Average synaptic weight $W[t] = \avg{W_{ij}[t]} $ as a function of time. Synaptic weights converge to the stationary state $W^*\approx W_c=0.45$ after perturbation.
    The network is periodically perturbed at every $t_p=400n$~ ms, $n=1,\cdots,8$, artificially setting $W_{ij}[t_p]=0.01(n+1)$.
    \textbf{B.} Stationary $W^*$ versus various initial conditions $W[0] = \avg{W_{ij}[t=0] }$, 
    $N =200, A = 0.06, \tau = 1000, U_W = 0.1$.
    Dashed lines are guides for the eyes to locate the quasi-critical region where $W\in [0.04,0.05]$.
    In this and in all the next figures the error bars are smaller than the symbols;
    \textbf{C.} Spontaneous activity $\rho[t]$ (black, upper) and average membrane potential $V[t]$
(red, lower) in the quasi-critical region: note the stochastic switching between large and small amplitude oscillations
due to homeostatic hovering around the synchronization transition ($A = 0.058$, $W^* \approx 0.4$, $N = 200$, $\tau=1000$, $U_W = 0.1$).}
\end{figure*}

The $\alpha$ and $\tau_A$ of the best fit are shown in Fig.~\ref{fig:reverbfit}C,
\adicionado{yielding} $\bar{\alpha}=0.20\pm0.05$ (mean$\pm$SD over all best fits, see Supplementary Information).
This temporal power-law decay of the activity is associated with
critical slowing down in equilibrium phase transitions~\cite{Cocchi2017}.
The damping time is optimized at $W=0.043$, close to $W_c$ and
within the critical range of the phase transition.

\subsection{Synchronization-avoiding homeostatic plasticity}

The input reverberation advantage at the critical point begs the question:
Is there a mechanism that is capable of keeping the network functioning around the synchronization critical point?
If so, what are the necessary ingredients for this dynamics without the need to fine-tune the synaptic couplings $W_{ij}$?

We look for a dynamic that is capable of homeostatically going around a synchronization
transition. Thus, if the network is synchronized, it must decrease the coincidence of spikes;
otherwise, it should allow spikes to happen freely.
Inspired by previous work~\cite{Brochini2016,Costa2018,Kinouchi2019,Menesse2021,Menesse2023}
\adicionado{(see Discussion)},
we then introduce the following \adicionado{anti-Hebbian} plasticity:
\begin{equation} 
\label{Wt}
W_{ij}[t+1] = W_{ij}[t] +\dfrac{1}{\tau}
\left(A -W_{ij}[t]\right) - U_W W_{ij} S_i[t] S_j[t] \:.
\end{equation}
\adicionado{This means that a synapse is depressed by a factor of $U_W W_{ij}$ whenever the
pre- and postsynaptic spikes coincide, $S_i[t]S_j[t]=1$. Otherwise,} the weight
$W_{ij}$ is restored to the baseline synaptic coupling strength $A$ on a time scale $\tau$.
\adicionado{Since synchronization increases spike coincidence, this mechanism
is expected to avoid it. It also uses only local information to change synapses,
since postsynaptic information $S_i[t]$
can be transmitted by retrograde dendritic spikes~\cite{Holthoff2006,Gollo2009}, similarly to
what happens for spike-timing dependent plasticity~\cite{Shimoura2021}.}

\adicionado{For this part of the study, we cannot assume homogeneous weights beforehand. Thus,
the initial couplings $W_{ij}[t=0]$ are chosen at random following a Gaussian
distribution with mean over off-diagonal pairs $\avg{W_{ij}[0]}\neq W_c$.
The variance of the initial distribution is not important, provided that all the $W_{ij}[0]$
are positive (excitatory).
For each time step, the average synaptic weight is $W[t]=\avg{W_{ij}[t]}$.
When reached, the steady state can be evaluated by taking $W^*= \avg{ \avg{W_{ij}}}_t$ (first over off-diagonal pairs, then over time).
Contrary to previous models, here the average synaptic coupling $W^*$ depends on the pairwise site correlation, $\avg{S_iS_j}$.
This means that the mechanism we propose does not require an absorbing state to be able to drive the system
towards the edge of the synchronization transition, thereby avoiding synchronization -- see Discussion.}

\adicionado{Recall that the synchronization phase transition starts at the critical point $W_c$, although the network is only fully synchronized when $\chi\sim1$ for
$W>W_c$. Thus, if our mechanism is capable of robustly converging to $W^*\sim W_c$ (avoiding $W$ values where $\chi\sim1$),
we can safely say that it is successfully avoiding synchronization.}
Such an asymptotic limit \adicionado{($W^*\sim W_c$)} should be robust and
depend only on the gross tuning of the hyperparameters ($A$, $U_W$, $\tau$),
regardless of initial conditions. \adicionado{We thus set out to investigate these features.}

\begin{figure*}[!htp]
    \centering    
    \includegraphics[width=\textwidth]{{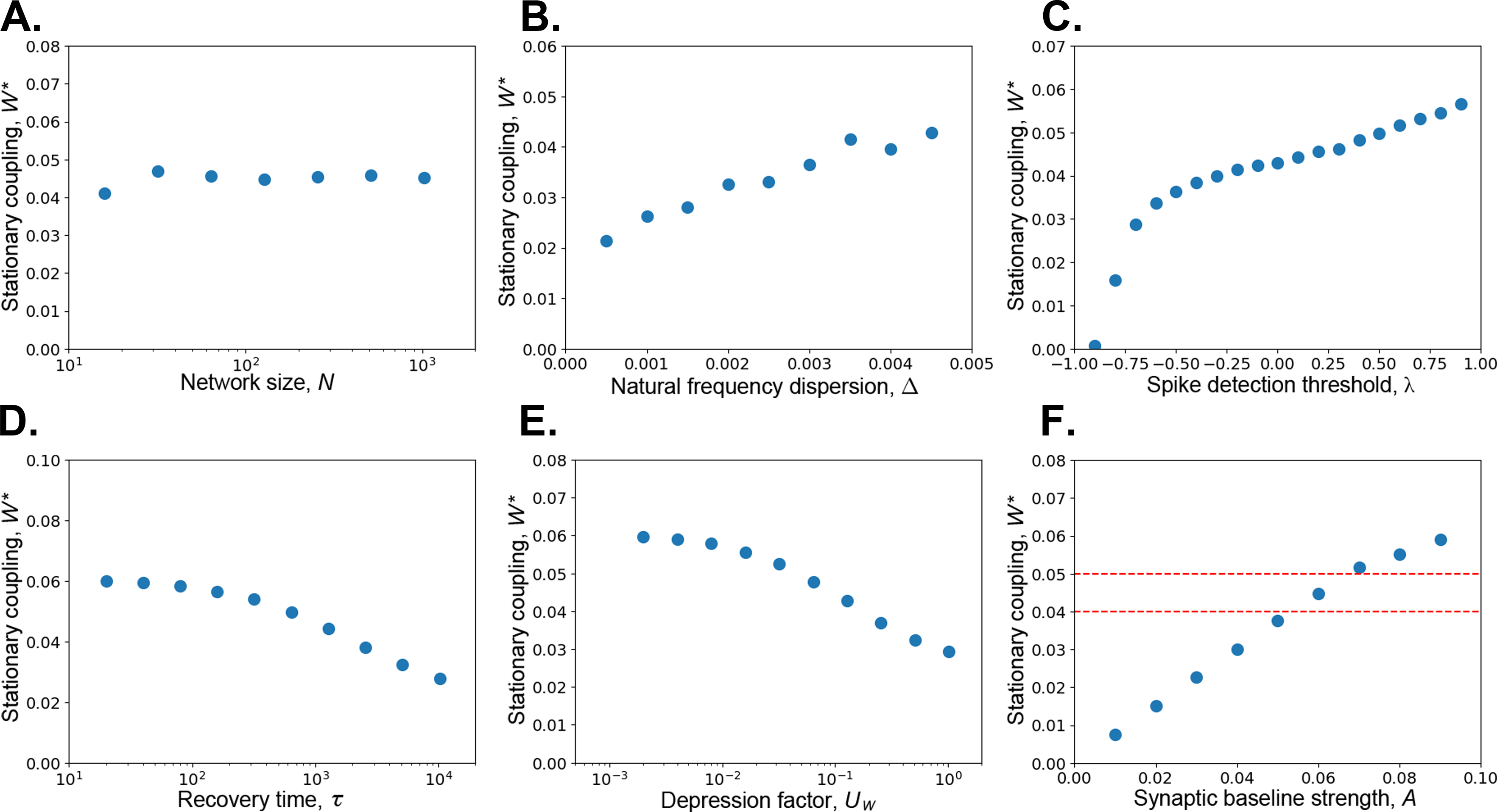}}
    \caption{\label{fig:hyperparam}
    \textbf{Conditions for reaching the critical point.} Each panel shows the behavior of the stationary state $W^*$
    as a function of a given parameter. All other parameters are kept fixed.
    \textbf{A.} The convergence to the critical point is robust for varying network size $N$. $A=0.06$, $\tau=1000$ and $U_W=0.1$.
    \textbf{B.} The dynamics tracks the critical that shifts with $\Delta$. $N=200$, $A=0.06$, $\tau=1000$ and $U_W=0.1$.
    \textbf{C.} The spike detection threshold $\lambda$ has a convenient value (around zero), but is just a proxy for some underlying spike detection mechanism. $N = 200, A = 0.06, \tau=1000$ and $U_W = 0.1$.
    \textbf{D.} The stationary state depends logarithmically on the depression recovery time $\tau$. $N=200$, $A=0.06$ and $U_W=0.1$.
    \textbf{E.} $W^*$ also depends logarithmically on $U_W$. $N=200$, $A=0.06$ and $\tau=1000$.
    \textbf{F.} The parameter $A$ ultimately determines $W^*$ and cannot be placed too far from $W_c$. $N=200$, $\tau=1000$ and $U_W = 0.1$.
    The dashed lines mark the critical region. The error bars are smaller than the symbols.
    }
\end{figure*}

\adicionado{In Fig.~\ref{fig:Wt}, we can see that} $W^*$ converges to near $W_c$, within the quasicritical region $0.04\leq W\leq0.05$.
This behavior is robust against temporal perturbations \adicionado{(Fig.~\ref{fig:Wt}A)}
and different initial conditions (Fig.~\ref{fig:Wt}B).
The steady-state activity shows homeostatic oscillations similar to those typically observed
in absorbing phase transition systems~\cite{Kinouchi2019,Girardi2020}. However,
here the system keeps switching between large and small amplitude oscillations (Fig.~\ref{fig:Wt}C).
\adicionado{Additionally, Fig.~\ref{fig:hyperparam}A shows that the dynamics is robust both for small and large systems.}

\subsubsection{Dependence on the natural frequency dispersion $\Delta$}

$\Delta$ defines the broadness of the natural frequency distribution of neurons.
The infinite-period bifurcation shown in the $\delta\times Q$ plane (Fig.~\ref{fig:phasediag})
suggests that $W_c$ (the synchronization point with constant $W_{ij}$) is a function of $\Delta$. 
Larger $\Delta$ results in increased $W_c$.
Introducing the homeostatic dynamics, we obtain growing $W^*$ with increasing $\Delta$ (Fig.~\ref{fig:hyperparam}B).
In other words, this means that the stationary state tracks the boundary of the phase transition,
converging to the critical point $W_c$ for each $\Delta$.
Note that all other parameters are fixed ($A$, $\tau$, $U_W$ and $\lambda$).

\subsubsection{Effect of the spike detection threshold $\lambda$}

The threshold $\lambda$ defines
the width of the spike that is effectively \textit{perceived} by the depression term in $W_{ij}$.
For a given neuron, a smaller $\lambda$ implies in a longer time satisfying the condition $V_i[t]>\lambda$,
yielding a longer depression.
The value of $\lambda$ is arbitrary, since we use it only as a simple way to define the coincidence between spikes.
A possible biological mechanism that could implement the same detection is beyond the scope of our work.

There is a large interval $\lambda \in [-0.5;0.5]$ where the stationary value $W^*$ weakly depends on $\lambda$ and goes to the phase transition range
$W^*\in[0.04;0.05]$ near $W_c\approx0.045$ (Fig.~\ref{fig:hyperparam}C).
We choose $\lambda=0$ given the symmetry of the hyperbolic tangent
that defines the membrane potential.


\subsubsection{Dependence on hyperparameters}

The synaptic recovery time $\tau$ controls the speed with which
$W^*$ returns to the baseline $A$ in the absence of coincident spikes.
This means that we have $W^* \approx A$ for small $\tau$.
For large $\tau$, depression accumulates more quickly and makes $W^*$ underestimate $W_c$
in finite simulation times (Fig~\ref{fig:hyperparam}D). We hypothesize that longer simulation times (of the order of many $\tau$)
could result in $W^*\sim W_c$ if $\delta_i$ are small enough leaving the neurons with very slow autonomous spiking.
However, the asymptotic behavior of the stationary state is $W^* \sim -\log(\tau)$, which means that
there is a wide range of $\tau$ that leads to the critical region around $W_c$.


The depression strength $U_W$ controls the effect of
coincident spikes on the synaptic couplings.
$U_W$ and $\tau$ could be replaced by a single parameter~\cite{Menesse2021},
since both represent competing time scales.
Therefore, the behavior of the stationary state as a function of $U_W$ is equivalent
to that as a function of $\tau$.
We have $W^* \approx A$ for small $U_W$, and large depression for large $U_W$ (Fig~\ref{fig:hyperparam}E),
underestimating $W_c$ in a finite simulation time.
Again, the asymptotic behavior of the system is $W^* \sim -\log(U_W)$, yielding a relatively long
regime of near-criticality along the $U_W$ range.


The hyperparameter
that has the greatest effect is the
synaptic baseline $A$.  This is because without depression, 
the dynamic continually pushes the
couplings toward $A$ like a restoring force,
resulting in $W^*=A$ exactly.

The depressing factor transforms this dynamic into sublinear with $A$ (Fig.~\ref{fig:hyperparam}F).
We cannot set $A$ too far from $W_c$, otherwise
the system would not converge to the critical point. Indeed, this has been known since the original
introduction of the depressing synapses~\cite{Levina2007}, and has been confirmed by a
series of studies in systems with absorbing phase
transitions~\cite{Brochini2016,Costa2018,Kinouchi2019,Girardi2020,Girardi-Schappo2021,Menesse2021}.
Such tuning is unavoidable for this type of homeostatic dynamics~\cite{Girardi-Schappo2021,Menesse2023}
(see Discussion).

\subsubsection{Input reverberation with homeostatic synchronization}

\adicionado{Now that we have studied the properties of the input reverberation in networks without adaptation, and understood the
role of hyperparameters in the synchronization dynamics, we can probe for the properties of the homeostatic networks.}

\begin{figure}[!tp]
    \centering    
    \includegraphics[width=0.4\textwidth]{{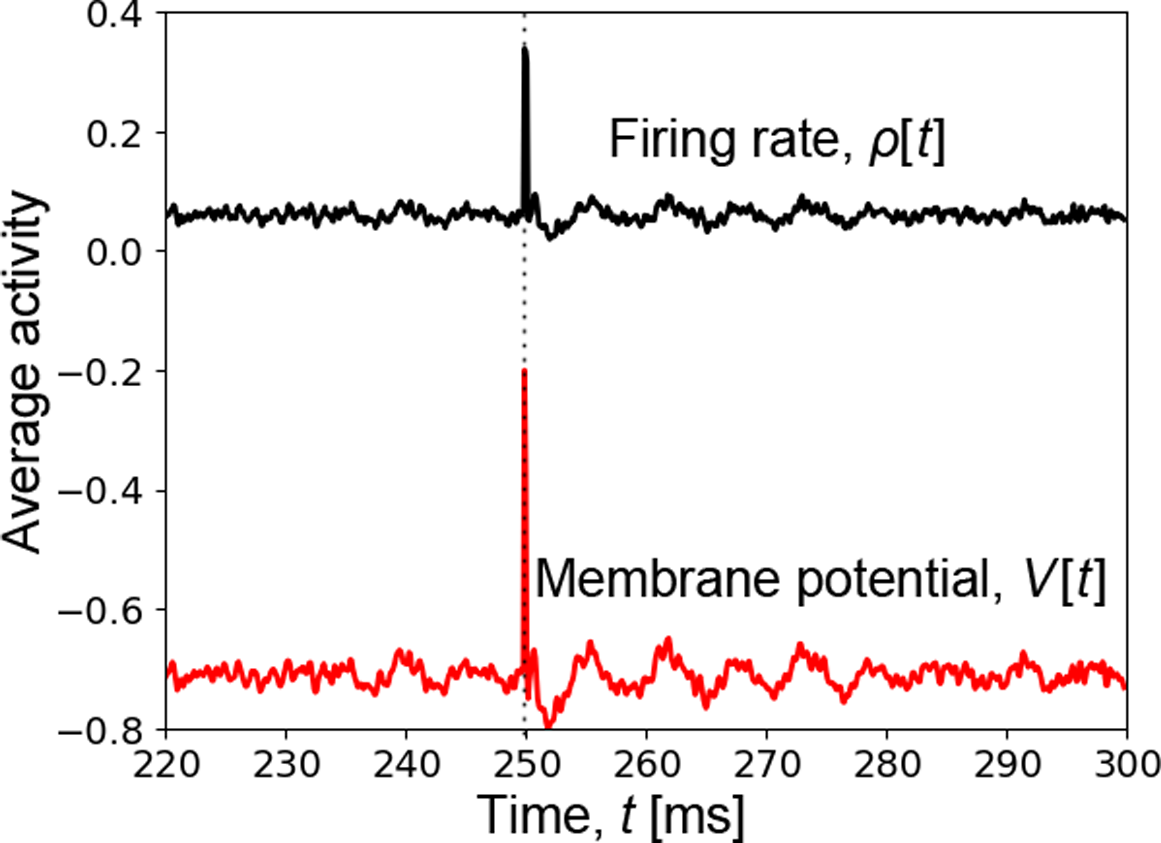}}
    \caption{\label{fig:SWA}
    \textbf{Input reverberation on a homeostatic network with dynamical synapses.}
    The transient oscillations are visible in the firing activity $\rho[t]$ (black, top) and in the
    average voltage $V[t]$ (red, bottom) after an stimulus of intensity $I_0=0.3$ (vertical dotted line). $A=0.058$, $W^*\approx0.045$, $N=1000$, $\Delta=0.003$, $\tau=1000$ and $U_W=0.1$.}
\end{figure}

These systems also show
input reverberation (Fig.~\ref{fig:SWA}), mimicking the near-critical behavior
of the static network with $W\approx W_c$.
This is somewhat counterintuitive, since inputs create instantaneous synchrony, which in turn generates spike
coincidence and strongly depresses synapses.

\section{Discussion}

We introduced a map-based neuron with naturally occurring spikes
and studied its bifurcations. Collectively adjusted in a tonic spiking mode and placed in an all-to-all network with diffusive couplings,
these neurons synchronize through a continuous phase transition (defining a critical point $W_c$ on the average coupling strengths).

\adicionado{The input reverberation duration is optimized at the transition point due to critical slowing down.
We hypothesize that this could optimize information processing, since transient oscillations could be used for computations~\cite{Brette2012,Palmigiano2017}.
In the subcritical region, synchrony is quickly lost and the network presents almost no transient
oscillations.
In the near-critical region and at the critical point, long
transient oscillations appear after a perturbation
(this is known as input reverberation) due to critical slowing down.
The supercritical (synchronous) phase already has oscillations; and after the stimulus,
the network quickly returns to this stationary state, yielding almost no transient dynamics.
The reverberating input could work as a memory buffer for a much longer period of time than in the subcritical or supercritical cases.}

Then, we proposed a self-organization mechanism that is capable of dynamically reaching the synchronization point and staying there.
The mechanism was inspired by previous work, with the fundamental difference that here the synaptic depression is caused
by the coincidence of pre and postsynaptic spikes. This generates homeostatic dynamics,
leading the system to duel close to the boundary of the synchronization phase transition.
The resulting adaptive network has the following properties:
a) an attractor (stationary) state $W^*$ appears for the average coupling;
b) this attractor is robust to temporal perturbations;
c) there is a logarithmically large range of the hyperparameters in which $W^*$ stays around the boundaries of the critical point $W_c$.
Recently, a noisy homeostasis near a synchronization transition was proposed to model the sensitivity of the snake pit organ~\cite{Graf2024Bif}.
Ideally, the baseline $A$ should be near the critical point that leans toward the supercritical region.
However, its precise value is not important since $W^*$ also depends on
the time scales $\tau$ and $U_W$.
This feature also occurs in all systems that have similar types of
adaptation~\cite{Levina2007,Bonachela2010,Costa2018,Kinouchi2019,Girardi2020,Girardi-Schappo2021,Menesse2021,Menesse2023}.

\adicionado{In recent experiments~\cite{van2018,Tal2020,Schmidt2023}, oscillatory transients \adicionado{similar to those in Fig.~\ref{fig:Wt}C}
(called ``oscillatory bursts'') have been observed in M/EEG signals.
We hypothesize that this could also be evidence of a synchronization-avoiding dynamics as proposed here,
where the system homeostatically goes around a synchronization critical point.
In absorbing systems with similar coupling dynamics, stochastic oscillations are driven by finite-size noise that perturbs a weakly stable
focus~\cite{Costa2017,Kinouchi2019}.
Here, the mechanism should be analogous, although the stable focus should be replaced by a stable spiral that is reminiscent of
a limit cycle (that appears in the network for $W>W_c$).}

\subsection{\adicionado{Why map-based neurons?}}

\adicionado{The neuron model we used offers a compromise between the dynamical properties of
conductance-based models, analytical tractability, and computational efficiency, showing a phenomenology analogous to the Hindmarsh-Rose model~\cite{Hindmarsh1984}.
Maps offer a full framework for complex dynamics, from chaotic bursting to adaptable spiking threshold, to stochastic resonance~\cite{Girardi2013,Girardi2017}.
Furthermore, our map is richer than two dimensional systems like FitzHugh-Nagumo neurons and coupled Kaneko oscillators~\cite{kaneko1994CML}. }

\adicionado{Maps bring new insights and new challenges to the modeling of the synchronization phenomenon. 
Differently from the IF neurons, the spike in our map is not instantaneous; it is generated from the interplay
between the $K$, $T$, $\delta_i$ and $u$ time scales. It has all the spike phases that are traditionally observed in conductance-based models:
up-rise and downfall, followed by a variable refractory period (after-hyperpolarizing potential), and it has no well-defined
firing threshold, making the spike a continuous process.
These features alone pose a big challenge to the modeling, since the information in the network is neither instantaneously nor certainly
transmitted through synaptic couplings. On the other hand, this allowed us to give an account of the necessary individual dynamics 
that can generate the collectively observed synchronization: here, the neurons need to have their slow recovery time scales ($\delta$)
distributed close to an infinite-period bifurcation.}

\adicionado{
The collective dynamics of map networks is not as trivially inferred from the microscopic details as in the case of IF networks (see, \textit{e.g.}, \cite{kaneko1993CMLbook}). 
For example, power-law neuronal avalanche are easily observed in IF networks, since they are models of branching processes~\cite{Carvalho2021} due to the
instantaneous, and hence discrete, nature of the spike.
Conversely, these avalanches require a delicate interplay between the synaptic and neuronal time scales in map-based neuronal networks; something that cannot be achieved with the
simple tuning of a single parameter~\cite{Girardi2013b}. Here, we took the time scale balance into account by choosing the values of $T$, $K$ and $H$ that give
rise to the homoclinic bifurcation underlying the synchronization transition.}

\adicionado{
More complex models like ours are required to either confirm the universality hypothesis of the phase transitions that are predicted by
simple models with stereotyped microscopic dynamics, such as the coupled Kuramoto oscillators;
or to introduce new nonequilibrium universality classes~\cite{kaneko1993CMLbook}.
We observed a critical exponent $\alpha=0.20\pm0.05$ for the amplitude decay over time, pointing to a collective bifurcation
different from the supercritical Neimark-Sacker. However, further studies are necessary to investigate the scaling law of the phase transition
in our model, although this is beyond the scope of the current work.
}

\subsection{To tune or not to tune}

As nicely discussed by \citeauthor{hernandez2017self}~\cite{hernandez2017self},
the tuning of some parameters is unavoidable in the study of self-organization.
Here, self-organization occurred because we turned the control parameter $W$ into a slowly changing dynamical quantity via negative feedback from
local information. This is enough to demonstrate the capacity of the system to reach the transition point autonomously, even though this will
depend on the new set of parameters that controls the feedback (which we called the hyperparameters).

Due to homeostatic dynamics, there is a finite region in the hyperparameter space ($\tau$, $U_W$, $A$) where
the system is capable of autonomously reaching the synchronization point $W_c$.
This region is relatively large, since $W^*$ depends on
$\log \tau$ and $\log U_W$.
Regardless of the initial distribution of the synaptic weights $P(W_{ij}, t=0)$, the weights
self-organize into a stationary $P^*(W_{ij})$ with average $W^*$ that can be as close to $W_c$ as desired.

Such a tuning of the parameters that govern homeostasis happens even in the standard sandpile model~\cite{Dickman2000}.
There, the accumulation and dissipation time scales have to be adjusted so that the excess of grains is controlled via feedback
with the avalanching dynamics.

\subsection{\adicionado{The absorbing state is not necessary for the Synchronization-Avoiding Self-Organization (SASO)}}

\adicionado{Several homeostatic mechanisms have been proposed to
self-organize a network to criticality; for a review,
see~\cite{Kinouchi2020}. They are based on the
same principle: If the network has small activity $\rho\sim0$ (\textit{i.e.}, if the system is in the absorbing phase $W<W_c$, and is subcritical),
then $W$ must slowly grow towards $W_c$; or else if the network activity is too large $\rho\approx1$ (the system is in the active phase,
$W>W_c$, and is supercritical), then $W$ must decrease towards $W_c$.
The same can be accomplished with parameters other than synaptic coupling~\cite{Kinouchi2020}.
This mechanism is asymptotically present even in the Sandpile model~\cite{Dickman2000} -- the seminal
model for Self-Organized Criticality (SOC)~\cite{Bak1987}.
In the context of neural networks, it was introduced by \citeauthor{Levina2007}~\cite{Levina2007} through short-term synaptic depression and recovery.
As a consequence, the activity of the network displays stochastic homeostatic oscillations around the critical point~\cite{Bonachela2010,Kinouchi2020}.
}

\begin{figure}[!tp]
    \centering    
    \includegraphics[width=0.4\textwidth]{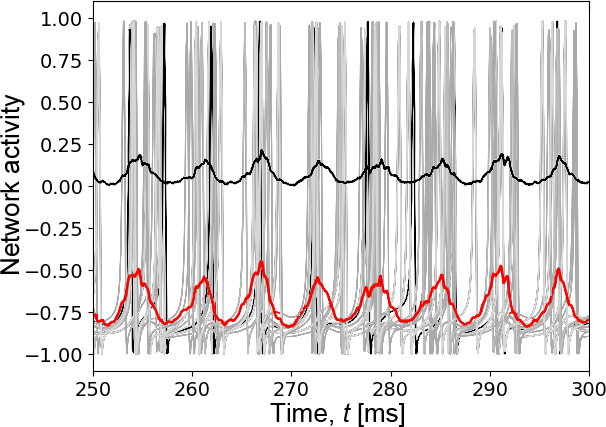}
    \caption{\label{fig:epi}
    \textbf{Synchronous \textit{seizure\adicionado{-like}} activity due to impaired homeostasis}.
    Example of the homeostatic network activity with an impaired combination of homeostatic parameters.
    In this case, the stationary state $W^*$ converges to the supercritical synchronous state.
    Average voltage $V = \avg{V_i}$ (red oscillations, bottom); average activity $\rho=\avg{S_i}$ (black oscillations above zero).
    Background: spikes of a few neurons selected at random.
    $N=1000$, $W^*\approx0.064$, $A=0.1$, $\tau=1000$, $U_W=0.1$.}
\end{figure}

\adicionado{SOC and its sister mechanisms (such as Self-Organized quasi-Criticality -- SOqC, Self-Organized Bistability -- SOB,
and Self-Organized Collective Oscillations -- SOCO) are theoretically interpreted via stochastic Langevin equations for the firing rate $\rho$ coupled
to an average adaptive synaptic field $W$~\cite{Buendia2020}. A common feature connecting the dynamics of all these theories is that the phase transition
must occur towards an absorbing state where the firing rate is $\rho=0$. 
The reason is simple: the microscopic dynamics that give rise to these theories have a synaptic depression term of the form $-U_WW_{ij}S_j$,
where $S_j=1$ when a presynaptic spike happens (zero otherwise) and $U_W>0$ is an arbitrary constant~\cite{Kinouchi2020,Buendia2020};
$W_{ij}$ is the synaptic coupling conductance.
Thus, the averaged synaptic coupling field is depressed proportionally to the firing rate, since $U_W\avg{W_{ij}S_j}=U_WW\rho$, and the depression
turns off whenever $\rho=0$ allowing the synaptic coupling to be restored.}

\adicionado{In contrast, the microscopic mechanism we proposed in Eq.~\eqref{Wt} depends on the coincidence of pre- and postsynaptic spikes,
which gives rise to an average adaptive synaptic field that is depressed proportionally to the pairwise site correlation, since
$U_W\avg{W_{ij}S_iS_j}=U_WW\avg{S_iS_j}$. The recovery of the average synaptic field starts only when $\avg{S_iS_j}=0$,
which is directly related to a null synchronization index $\chi=0$.
This is because $\chi^2$ is the variance of the network relative to the individual membrane potentials.
Notwithstanding, $\chi$ is our order parameter of choice.
We name this mechanism by Synchronization-Avoiding Self-Organization, or SASO, and it works even in the absence
of an absorbing state. In other words,
the system may have non-vanishing firing rate in both sides of the critical point $W_c$,
and synchronization will still be prevented if the conditions that we unveiled for the hyperparameters are obeyed.}

\subsection{Relevance to epilepsy}

Epilepsy is a complex phenomenon that depends on multiple factors~\cite{jiruska2013synchronization}.
Its expression depends on both the structural~\cite{Bernhardt2013} and the dynamical~\cite{Girardi2021Epil} features.
In the early observation of neuronal avalanches, seizures were hypothesized to correspond to a supercritical state with synchronized spikes
and oscillations at the average population voltage~\cite{Beggs2003}.
Other authors have also suggested that epilepsy is related to loss of criticality~\cite{Meisel2012,Meisel2020EpiSub,Maturana2020}, but no mechanism has been proposed.
A synchronization transition can be used as a simple model for epileptic seizures~\cite{jiruska2013synchronization,Rich2020Epi}.
In this context, there are some attempts to model the transition into ictal activity (\textit{e.g.}, fast oscillations with large amplitude)
both at the population~\cite{Rich2020Epi} and at the electroencephalography~\cite{jirsaEpileptor2014} levels.
However, only qualitative comparisons have been made between self-organized critical models and empirical data~\cite{Meisel2012}.
\adicionado{Thus, considering the similarities between the dynamics in our model and the synchronization feature of seizure-like
activity observed experimentally~\cite{Ziburkus2013EIseizure,Chen2001seizure,Naylor2010seizure},
we study the transition to synchronization as a theoretical approximation of the onset of seizure-like activity.}

Some authors suggest that the ictal state may be elicited by
increased excitation resulting from an underlying neuromodulatory process~\cite{LopesdaSilva2003}.
In our model, this is captured by combining the parameters of synaptic depression in such a way that $W^*\approx A$,
disrupting the near-criticality homeostatic balance\adicionado{, and driving the system towards supercritical synchronized oscillations.}
Namely, considering $A>W_c$ some malfunction could produce a combination of the recovery time $\tau$ and depression factor $U_W$ that is
incapable of suppressing $W^*$ sufficiently from the baseline $A$.
This \adicionado{would keep} the network hyperexcited, favoring synchronization and generating fast and large waves (Fig.~\ref{fig:epi}).
An analogous transition is achieved by an intricate interplay of slow-fast time scales
in \adicionado{another} phenomenological model for the EEG activity of the brain~\cite{jirsaEpileptor2014}.

Reflex seizures can also be triggered by external stimuli that statistically favor synchrony~\cite{Trenite2011,Hermes2017,Honey2017}.
This is also captured by our model:
Being on the verge of the synchronization phase transition, our network is capable of reverberating synchrony-provoking stimuli over long times (Fig.~\ref{fig:SWA}).
Although this \adicionado{could sometimes be} advantageous to computation~\cite{Izhikevich2006,Brette2012},
a faulty configuration of neuronal and synaptic time scales (governed by $T$, $\delta_i$, $K$, $\tau$ and $U_W$)
\adicionado{could also} excessively delay the damping of the response oscillations beyond the \adicionado{optimal $\tau_A$ value}.
\adicionado{Consequently, the loss of damping would turn the input reverberation into sustained
supercritical synchronized oscillations, triggering} seizure\adicionado{-like activity}.

\subsection{\adicionado{Other attempts to suppress synchronization}}

\adicionado{
Some studies have looked at the influence of synaptic plasticity on synchronization~\cite{sayari2022},
or found first order synchronization transitions in map-based networks~\cite{boaretto2019chialvo}.
A theory has also been recently proposed~\cite{ClarkAbbott2024RateAdapt} to describe Hebbian and anti-Hebbian synaptic adaptation in
Sompolinsky \textit{et al.}~\cite{Sompolinsky1988Rate}-like rate networks.
Here, we proposed a simulation of a spiking neural networks that can be explained under the umbrella of the SOC-derived theories~\cite{Buendia2020},
where the adaptative field receives feedback from the pairwise site correlation to avoid synchronization.
Others have developed a global mechanism that can be used to control the synchronization near a Hopf bifurcation~\cite{Rosenblum2004Feedback}.
Controlling or even suppressing synchronization is a timely subject, given that oscillations are ubiquitously
found both in nature and in many engineering applications.}

\adicionado{
In the context of the brain, synchronization-suppressing control feedback could be used to
extinguish pathological oscillations. The method consists in adding a direct
feedback of the average membrane potential onto itself after a given delay~\cite{Rosenblum2004Feedback}.
The method was successfully applied to mean-field networks of Hindmarsh-Rose and Rulkov neurons.
This strategy was also successful in eliminating the synchronization of hierarchically clustered complex networks
of bursting Rulkov maps coupled by different types of synapses~\cite{lameu2016suppression,mugnaine2018delayed,reis2021bursting}.
}

\adicionado{The main difference is that ours is a local mechanism, since it involves only the pre- and postsynaptic spikes.
Not only that, but in the feedback control theory, the feedback weight that couples the global average membrane potential
to each neuron is constant. Conversely, in our mechanism the coupling constant is itself adaptive, and it depends only on local information.
Our mechanism is not intended to be used as a form of externally controlling the synchronization. Rather, we hypothesize
that it could be an intrinsic mechanism built-in to the network in order to avoid pathological synchronization.}


\section{Perspectives}

The simplified dense topology allowed us to isolate the effects of the dynamics on the synchronization.
Therefore, the effect of connectivity is not yet explored.
Our model can also be expanded by considering chemical synapses
and excitatory and inhibitory populations of neurons.
The role of synaptic noise in the system and how to regulate the hyperparameters of the homeostasis
are also open questions.

Our findings not only contribute to the broader understanding of brain criticality,
but also present innovative perspectives on the role of synchronization phase
transitions and critical slowing down in brain information processing. The
proposed homeostatic synaptic dynamics adds a novel dimension to the exploration
of network behavior near bifurcation points and yields simplified (yet general) explanations for neurological disorders.

\section*{Supplementary Material}

We provide extra figures and information about the microscopic activity of the network and the fitting of the damping characteristic time.

\begin{acknowledgments}
S.L.R. acknowledges a CAPES Ph.D. fellowship. O.K. acknowledges CNAIPS-USP, FAPESP support, and
a CNPq research fellowship. S.L.R. and O.K. thank the
CEPID NEUROMAT support. The authors are grateful for advice from
Mauro Copelli and Antônio C. Roque.
\end{acknowledgments}

\section*{Author Declarations}
\subsection*{Conflict of Interest}
The authors have no conflicts to disclose.

\subsection*{Author Contributions}
S.L.R.: Conceptualization (equal); Investigation (equal); Validation (equal); Visualization (equal); Writing the original draft (equal).
M.G.-S.: Investigation (equal); Validation (equal); Supervision (equal); Visualization (equal); Writing (equal).
O.K.: Conceptualization (equal);  Investigation (equal); Supervision (equal); Validation (equal); Writing original draft (equal).

\section*{Data Availability Statement}

The data that support the findings of this study are available within the article and its supplementary material.

\section*{References}

\begin{thebibliography}{85}%
\makeatletter
\providecommand \@ifxundefined [1]{%
 \@ifx{#1\undefined}
}%
\providecommand \@ifnum [1]{%
 \ifnum #1\expandafter \@firstoftwo
 \else \expandafter \@secondoftwo
 \fi
}%
\providecommand \@ifx [1]{%
 \ifx #1\expandafter \@firstoftwo
 \else \expandafter \@secondoftwo
 \fi
}%
\providecommand \natexlab [1]{#1}%
\providecommand \enquote  [1]{``#1''}%
\providecommand \bibnamefont  [1]{#1}%
\providecommand \bibfnamefont [1]{#1}%
\providecommand \citenamefont [1]{#1}%
\providecommand \href@noop [0]{\@secondoftwo}%
\providecommand \href [0]{\begingroup \@sanitize@url \@href}%
\providecommand \@href[1]{\@@startlink{#1}\@@href}%
\providecommand \@@href[1]{\endgroup#1\@@endlink}%
\providecommand \@sanitize@url [0]{\catcode `\\12\catcode `\$12\catcode
  `\&12\catcode `\#12\catcode `\^12\catcode `\_12\catcode `\%12\relax}%
\providecommand \@@startlink[1]{}%
\providecommand \@@endlink[0]{}%
\providecommand \url  [0]{\begingroup\@sanitize@url \@url }%
\providecommand \@url [1]{\endgroup\@href {#1}{\urlprefix }}%
\providecommand \urlprefix  [0]{URL }%
\providecommand \Eprint [0]{\href }%
\providecommand \doibase [0]{http://dx.doi.org/}%
\providecommand \selectlanguage [0]{\@gobble}%
\providecommand \bibinfo  [0]{\@secondoftwo}%
\providecommand \bibfield  [0]{\@secondoftwo}%
\providecommand \translation [1]{[#1]}%
\providecommand \BibitemOpen [0]{}%
\providecommand \bibitemStop [0]{}%
\providecommand \bibitemNoStop [0]{.\EOS\space}%
\providecommand \EOS [0]{\spacefactor3000\relax}%
\providecommand \BibitemShut  [1]{\csname bibitem#1\endcsname}%
\let\auto@bib@innerbib\@empty
\bibitem [{\citenamefont {Brette}(2012)}]{Brette2012}%
  \BibitemOpen
  \bibfield  {author} {\bibinfo {author} {\bibfnamefont {R.}~\bibnamefont
  {Brette}},\ }\bibfield  {title} {\enquote {\bibinfo {title} {Computing with
  neural synchrony},}\ }\href {\doibase 10.1371/journal.pcbi.1002561}
  {\bibfield  {journal} {\bibinfo  {journal} {PLoS Comput Biol}\ }\textbf
  {\bibinfo {volume} {8}},\ \bibinfo {pages} {e1002561} (\bibinfo {year}
  {2012})}\BibitemShut {NoStop}%
\bibitem [{\citenamefont {Izhikevich}(2006)}]{Izhikevich2006}%
  \BibitemOpen
  \bibfield  {author} {\bibinfo {author} {\bibfnamefont {E.~M.}\ \bibnamefont
  {Izhikevich}},\ }\bibfield  {title} {\enquote {\bibinfo {title}
  {{Polychronization: Computation with Spikes}},}\ }\href {\doibase
  10.1162/089976606775093882} {\bibfield  {journal} {\bibinfo  {journal}
  {Neural Comput.}\ }\textbf {\bibinfo {volume} {18}},\ \bibinfo {pages}
  {245--282} (\bibinfo {year} {2006})}\BibitemShut {NoStop}%
\bibitem [{\citenamefont {Palmigiano}\ \emph {et~al.}(2017)\citenamefont
  {Palmigiano}, \citenamefont {Geisel}, \citenamefont {Wolf},\ and\
  \citenamefont {Battaglia}}]{Palmigiano2017}%
  \BibitemOpen
  \bibfield  {author} {\bibinfo {author} {\bibfnamefont {A.}~\bibnamefont
  {Palmigiano}}, \bibinfo {author} {\bibfnamefont {T.}~\bibnamefont {Geisel}},
  \bibinfo {author} {\bibfnamefont {F.}~\bibnamefont {Wolf}}, \ and\ \bibinfo
  {author} {\bibfnamefont {D.}~\bibnamefont {Battaglia}},\ }\bibfield  {title}
  {\enquote {\bibinfo {title} {Flexible information routing by transient
  synchrony},}\ }\href@noop {} {\bibfield  {journal} {\bibinfo  {journal} {Nat
  Neurosci.}\ }\textbf {\bibinfo {volume} {20}},\ \bibinfo {pages} {1014--1022}
  (\bibinfo {year} {2017})}\BibitemShut {NoStop}%
\bibitem [{\citenamefont {Lehnertz}\ \emph {et~al.}(2009)\citenamefont
  {Lehnertz}, \citenamefont {Bialonski}, \citenamefont {Horstmann},
  \citenamefont {Krug}, \citenamefont {Rothkegel}, \citenamefont {Staniek},\
  and\ \citenamefont {Wagner}}]{Lehnertz2009Epi}%
  \BibitemOpen
  \bibfield  {author} {\bibinfo {author} {\bibfnamefont {K.}~\bibnamefont
  {Lehnertz}}, \bibinfo {author} {\bibfnamefont {S.}~\bibnamefont {Bialonski}},
  \bibinfo {author} {\bibfnamefont {M.-T.}\ \bibnamefont {Horstmann}}, \bibinfo
  {author} {\bibfnamefont {D.}~\bibnamefont {Krug}}, \bibinfo {author}
  {\bibfnamefont {A.}~\bibnamefont {Rothkegel}}, \bibinfo {author}
  {\bibfnamefont {M.}~\bibnamefont {Staniek}}, \ and\ \bibinfo {author}
  {\bibfnamefont {T.}~\bibnamefont {Wagner}},\ }\bibfield  {title} {\enquote
  {\bibinfo {title} {Synchronization phenomena in human epileptic brain
  networks},}\ }\href {\doibase https://doi.org/10.1016/j.jneumeth.2009.05.015}
  {\bibfield  {journal} {\bibinfo  {journal} {J Neurosci Methods.}\ }\textbf
  {\bibinfo {volume} {183}},\ \bibinfo {pages} {42--48} (\bibinfo {year}
  {2009})}\BibitemShut {NoStop}%
\bibitem [{\citenamefont {Rich}\ \emph {et~al.}(2020)\citenamefont {Rich},
  \citenamefont {Hutt}, \citenamefont {Skinner}, \citenamefont {Valiante},\
  and\ \citenamefont {Lefebvre}}]{Rich2020Epi}%
  \BibitemOpen
  \bibfield  {author} {\bibinfo {author} {\bibfnamefont {S.}~\bibnamefont
  {Rich}}, \bibinfo {author} {\bibfnamefont {A.}~\bibnamefont {Hutt}}, \bibinfo
  {author} {\bibfnamefont {F.~K.}\ \bibnamefont {Skinner}}, \bibinfo {author}
  {\bibfnamefont {T.~A.}\ \bibnamefont {Valiante}}, \ and\ \bibinfo {author}
  {\bibfnamefont {J.}~\bibnamefont {Lefebvre}},\ }\bibfield  {title} {\enquote
  {\bibinfo {title} {Neurostimulation stabilizes spiking neural networks by
  disrupting seizure-like oscillatory transitions},}\ }\href {\doibase
  10.1038/s41598-020-72335-6} {\bibfield  {journal} {\bibinfo  {journal} {Sci
  Rep}\ }\textbf {\bibinfo {volume} {10}},\ \bibinfo {pages} {15408} (\bibinfo
  {year} {2020})}\BibitemShut {NoStop}%
\bibitem [{\citenamefont {Jirsa}\ \emph {et~al.}(2014)\citenamefont {Jirsa},
  \citenamefont {Stacey}, \citenamefont {Quilichini}, \citenamefont {Ivanov},\
  and\ \citenamefont {Bernard}}]{jirsaEpileptor2014}%
  \BibitemOpen
  \bibfield  {author} {\bibinfo {author} {\bibfnamefont {V.~K.}\ \bibnamefont
  {Jirsa}}, \bibinfo {author} {\bibfnamefont {W.~C.}\ \bibnamefont {Stacey}},
  \bibinfo {author} {\bibfnamefont {P.~P.}\ \bibnamefont {Quilichini}},
  \bibinfo {author} {\bibfnamefont {A.~I.}\ \bibnamefont {Ivanov}}, \ and\
  \bibinfo {author} {\bibfnamefont {C.}~\bibnamefont {Bernard}},\ }\bibfield
  {title} {\enquote {\bibinfo {title} {On the nature of seizure dynamics},}\
  }\href {\doibase 10.1093/brain/awu133} {\bibfield  {journal} {\bibinfo
  {journal} {Brain}\ }\textbf {\bibinfo {volume} {137}},\ \bibinfo {pages}
  {2210--2230} (\bibinfo {year} {2014})}\BibitemShut {NoStop}%
\bibitem [{\citenamefont {Lopes~da Silva}\ \emph {et~al.}(2003)\citenamefont
  {Lopes~da Silva}, \citenamefont {Blanes}, \citenamefont {Kalitzin},
  \citenamefont {Parra}, \citenamefont {Suffczynski},\ and\ \citenamefont
  {Velis}}]{LopesdaSilva2003}%
  \BibitemOpen
  \bibfield  {author} {\bibinfo {author} {\bibfnamefont {F.}~\bibnamefont
  {Lopes~da Silva}}, \bibinfo {author} {\bibfnamefont {W.}~\bibnamefont
  {Blanes}}, \bibinfo {author} {\bibfnamefont {S.~N.}\ \bibnamefont
  {Kalitzin}}, \bibinfo {author} {\bibfnamefont {J.}~\bibnamefont {Parra}},
  \bibinfo {author} {\bibfnamefont {P.}~\bibnamefont {Suffczynski}}, \ and\
  \bibinfo {author} {\bibfnamefont {D.~N.}\ \bibnamefont {Velis}},\ }\bibfield
  {title} {\enquote {\bibinfo {title} {Epilepsies as dynamical diseases of
  brain systems: basic models of the transition between normal and epileptic
  activity},}\ }\href {\doibase 10.1111/j.0013-9580.2003.12005.x} {\bibfield
  {journal} {\bibinfo  {journal} {Epilepsia}\ }\textbf {\bibinfo {volume} {44
  Suppl 12}},\ \bibinfo {pages} {72--83} (\bibinfo {year} {2003})}\BibitemShut
  {NoStop}%
\bibitem [{\citenamefont {Beggs}\ and\ \citenamefont
  {Plenz}(2003)}]{Beggs2003}%
  \BibitemOpen
  \bibfield  {author} {\bibinfo {author} {\bibfnamefont {J.~M.}\ \bibnamefont
  {Beggs}}\ and\ \bibinfo {author} {\bibfnamefont {D.}~\bibnamefont {Plenz}},\
  }\bibfield  {title} {\enquote {\bibinfo {title} {Neuronal avalanches in
  neocortical circuits},}\ }\href@noop {} {\bibfield  {journal} {\bibinfo
  {journal} {J. Neurosci.}\ }\textbf {\bibinfo {volume} {23}},\ \bibinfo
  {pages} {11167--11177} (\bibinfo {year} {2003})}\BibitemShut {NoStop}%
\bibitem [{\citenamefont {Kinouchi}\ and\ \citenamefont
  {Copelli}(2006)}]{Kinouchi2006}%
  \BibitemOpen
  \bibfield  {author} {\bibinfo {author} {\bibfnamefont {O.}~\bibnamefont
  {Kinouchi}}\ and\ \bibinfo {author} {\bibfnamefont {M.}~\bibnamefont
  {Copelli}},\ }\bibfield  {title} {\enquote {\bibinfo {title} {Optimal
  dynamical range of excitable networks at criticality},}\ }\href {\doibase
  https://doi.org/10.1038/nphys289} {\bibfield  {journal} {\bibinfo  {journal}
  {Nat. Phys.}\ }\textbf {\bibinfo {volume} {2}},\ \bibinfo {pages} {348–351}
  (\bibinfo {year} {2006})}\BibitemShut {NoStop}%
\bibitem [{\citenamefont {Beggs}(2008)}]{Beggs2008}%
  \BibitemOpen
  \bibfield  {author} {\bibinfo {author} {\bibfnamefont {J.~M.}\ \bibnamefont
  {Beggs}},\ }\bibfield  {title} {\enquote {\bibinfo {title} {The criticality
  hypothesis: how local cortical networks might optimize information
  processing},}\ }\href@noop {} {\bibfield  {journal} {\bibinfo  {journal}
  {Philos. Trans. R. Soc. A}\ }\textbf {\bibinfo {volume} {366}},\ \bibinfo
  {pages} {329--343} (\bibinfo {year} {2008})}\BibitemShut {NoStop}%
\bibitem [{\citenamefont {Shew}\ and\ \citenamefont {Plenz}(2013)}]{Shew2013}%
  \BibitemOpen
  \bibfield  {author} {\bibinfo {author} {\bibfnamefont {W.~L.}\ \bibnamefont
  {Shew}}\ and\ \bibinfo {author} {\bibfnamefont {D.}~\bibnamefont {Plenz}},\
  }\bibfield  {title} {\enquote {\bibinfo {title} {The functional benefits of
  criticality in the cortex},}\ }\href@noop {} {\bibfield  {journal} {\bibinfo
  {journal} {Neuroscientist}\ }\textbf {\bibinfo {volume} {19}},\ \bibinfo
  {pages} {88--100} (\bibinfo {year} {2013})}\BibitemShut {NoStop}%
\bibitem [{\citenamefont {Cocchi}\ \emph {et~al.}(2017)\citenamefont {Cocchi},
  \citenamefont {Gollo}, \citenamefont {Zalesky},\ and\ \citenamefont
  {Breakspear}}]{Cocchi2017}%
  \BibitemOpen
  \bibfield  {author} {\bibinfo {author} {\bibfnamefont {L.}~\bibnamefont
  {Cocchi}}, \bibinfo {author} {\bibfnamefont {L.~L.}\ \bibnamefont {Gollo}},
  \bibinfo {author} {\bibfnamefont {A.}~\bibnamefont {Zalesky}}, \ and\
  \bibinfo {author} {\bibfnamefont {M.}~\bibnamefont {Breakspear}},\ }\bibfield
   {title} {\enquote {\bibinfo {title} {Criticality in the brain: A synthesis
  of neurobiology, models and cognition},}\ }\href {\doibase
  10.1016/j.pneurobio.2017.07.002} {\bibfield  {journal} {\bibinfo  {journal}
  {Prog. Neurobiol.}\ }\textbf {\bibinfo {volume} {158}},\ \bibinfo {pages}
  {132--152} (\bibinfo {year} {2017})}\BibitemShut {NoStop}%
\bibitem [{\citenamefont {Girardi-Schappo}(2021)}]{Girardi2021}%
  \BibitemOpen
  \bibfield  {author} {\bibinfo {author} {\bibfnamefont {M.}~\bibnamefont
  {Girardi-Schappo}},\ }\bibfield  {title} {\enquote {\bibinfo {title} {Brain
  criticality beyond avalanches: open problems and how to approach them},}\
  }\href {\doibase 10.1088/2632-072X/ac2071} {\bibfield  {journal} {\bibinfo
  {journal} {J Phys Complex}\ }\textbf {\bibinfo {volume} {2}},\ \bibinfo
  {pages} {031003} (\bibinfo {year} {2021})}\BibitemShut {NoStop}%
\bibitem [{\citenamefont {Plenz}\ \emph {et~al.}(2021)\citenamefont {Plenz},
  \citenamefont {Ribeiro}, \citenamefont {Miller}, \citenamefont {Kells},
  \citenamefont {Vakili},\ and\ \citenamefont {Capek}}]{plenz2021self}%
  \BibitemOpen
  \bibfield  {author} {\bibinfo {author} {\bibfnamefont {D.}~\bibnamefont
  {Plenz}}, \bibinfo {author} {\bibfnamefont {T.~L.}\ \bibnamefont {Ribeiro}},
  \bibinfo {author} {\bibfnamefont {S.~R.}\ \bibnamefont {Miller}}, \bibinfo
  {author} {\bibfnamefont {P.~A.}\ \bibnamefont {Kells}}, \bibinfo {author}
  {\bibfnamefont {A.}~\bibnamefont {Vakili}}, \ and\ \bibinfo {author}
  {\bibfnamefont {E.~L.}\ \bibnamefont {Capek}},\ }\bibfield  {title} {\enquote
  {\bibinfo {title} {Self-organized criticality in the brain},}\ }\href@noop {}
  {\bibfield  {journal} {\bibinfo  {journal} {Front. Phys.}\ }\textbf {\bibinfo
  {volume} {9}},\ \bibinfo {pages} {639389} (\bibinfo {year}
  {2021})}\BibitemShut {NoStop}%
\bibitem [{\citenamefont {O'Byrne}\ and\ \citenamefont
  {Jerbi}(2022)}]{Obyrne2022}%
  \BibitemOpen
  \bibfield  {author} {\bibinfo {author} {\bibfnamefont {J.}~\bibnamefont
  {O'Byrne}}\ and\ \bibinfo {author} {\bibfnamefont {K.}~\bibnamefont
  {Jerbi}},\ }\bibfield  {title} {\enquote {\bibinfo {title} {How critical is
  brain criticality?}}\ }\href@noop {} {\bibfield  {journal} {\bibinfo
  {journal} {Trends Neurosci.}\ }\textbf {\bibinfo {volume} {45}},\ \bibinfo
  {pages} {P820--837} (\bibinfo {year} {2022})}\BibitemShut {NoStop}%
\bibitem [{\citenamefont {Carvalho}\ \emph {et~al.}(2021)\citenamefont
  {Carvalho}, \citenamefont {Fontenele}, \citenamefont {Girardi-Schappo},
  \citenamefont {Feliciano}, \citenamefont {Aguiar}, \citenamefont {Silva},
  \citenamefont {de~Vasconcelos}, \citenamefont {Carelli},\ and\ \citenamefont
  {Copelli}}]{Carvalho2021}%
  \BibitemOpen
  \bibfield  {author} {\bibinfo {author} {\bibfnamefont {T.~T.~A.}\
  \bibnamefont {Carvalho}}, \bibinfo {author} {\bibfnamefont {A.~J.}\
  \bibnamefont {Fontenele}}, \bibinfo {author} {\bibfnamefont {M.}~\bibnamefont
  {Girardi-Schappo}}, \bibinfo {author} {\bibfnamefont {T.}~\bibnamefont
  {Feliciano}}, \bibinfo {author} {\bibfnamefont {L.~A.~A.}\ \bibnamefont
  {Aguiar}}, \bibinfo {author} {\bibfnamefont {T.~P.~L.}\ \bibnamefont
  {Silva}}, \bibinfo {author} {\bibfnamefont {N.~A.~P.}\ \bibnamefont
  {de~Vasconcelos}}, \bibinfo {author} {\bibfnamefont {P.~V.}\ \bibnamefont
  {Carelli}}, \ and\ \bibinfo {author} {\bibfnamefont {M.}~\bibnamefont
  {Copelli}},\ }\bibfield  {title} {\enquote {\bibinfo {title} {Subsampled
  directed-percolation models explain scaling relations experimentally observed
  in the brain},}\ }\href {\doibase 10.3389/fncir.2020.576727} {\bibfield
  {journal} {\bibinfo  {journal} {Front. Neural Circuits}\ }\textbf {\bibinfo
  {volume} {14}},\ \bibinfo {pages} {83} (\bibinfo {year} {2021})}\BibitemShut
  {NoStop}%
\bibitem [{\citenamefont {Ponce-Alvarez}\ \emph {et~al.}(2018)\citenamefont
  {Ponce-Alvarez}, \citenamefont {Jouary}, \citenamefont {Privat},
  \citenamefont {Deco},\ and\ \citenamefont {Sumbre}}]{PonceDeco2018zfish}%
  \BibitemOpen
  \bibfield  {author} {\bibinfo {author} {\bibfnamefont {A.}~\bibnamefont
  {Ponce-Alvarez}}, \bibinfo {author} {\bibfnamefont {A.}~\bibnamefont
  {Jouary}}, \bibinfo {author} {\bibfnamefont {M.}~\bibnamefont {Privat}},
  \bibinfo {author} {\bibfnamefont {G.}~\bibnamefont {Deco}}, \ and\ \bibinfo
  {author} {\bibfnamefont {G.}~\bibnamefont {Sumbre}},\ }\bibfield  {title}
  {\enquote {\bibinfo {title} {Whole-brain neuronal activity displays crackling
  noise dynamics},}\ }\href {\doibase 10.1016/j.neuron.2018.10.045} {\bibfield
  {journal} {\bibinfo  {journal} {Neuron}\ }\textbf {\bibinfo {volume} {100}},\
  \bibinfo {pages} {1446--1459.e6} (\bibinfo {year} {2018})}\BibitemShut
  {NoStop}%
\bibitem [{\citenamefont {Levina}, \citenamefont {Herrmann},\ and\
  \citenamefont {Geisel}(2007)}]{Levina2007}%
  \BibitemOpen
  \bibfield  {author} {\bibinfo {author} {\bibfnamefont {A.}~\bibnamefont
  {Levina}}, \bibinfo {author} {\bibfnamefont {J.~M.}\ \bibnamefont
  {Herrmann}}, \ and\ \bibinfo {author} {\bibfnamefont {T.}~\bibnamefont
  {Geisel}},\ }\bibfield  {title} {\enquote {\bibinfo {title} {Dynamical
  synapses causing self-organized criticality in neural networks},}\
  }\href@noop {} {\bibfield  {journal} {\bibinfo  {journal} {Nat. Phys.}\
  }\textbf {\bibinfo {volume} {3}},\ \bibinfo {pages} {857--860} (\bibinfo
  {year} {2007})}\BibitemShut {NoStop}%
\bibitem [{\citenamefont {Kinouchi}\ \emph {et~al.}(2019)\citenamefont
  {Kinouchi}, \citenamefont {Brochini}, \citenamefont {Costa}, \citenamefont
  {F},\ and\ \citenamefont {Copelli}}]{Kinouchi2019}%
  \BibitemOpen
  \bibfield  {author} {\bibinfo {author} {\bibfnamefont {O.}~\bibnamefont
  {Kinouchi}}, \bibinfo {author} {\bibfnamefont {L.}~\bibnamefont {Brochini}},
  \bibinfo {author} {\bibfnamefont {A.~A.}\ \bibnamefont {Costa}}, \bibinfo
  {author} {\bibfnamefont {C.~J.~G.}\ \bibnamefont {F}}, \ and\ \bibinfo
  {author} {\bibfnamefont {M.}~\bibnamefont {Copelli}},\ }\bibfield  {title}
  {\enquote {\bibinfo {title} {Stochastic oscillations and dragon king
  avalanches in self-organized quasi-critical systems},}\ }\href@noop {}
  {\bibfield  {journal} {\bibinfo  {journal} {Sci. Rep.}\ }\textbf {\bibinfo
  {volume} {9}},\ \bibinfo {pages} {3874} (\bibinfo {year} {2019})}\BibitemShut
  {NoStop}%
\bibitem [{\citenamefont {Girardi-Schappo}\ \emph {et~al.}(2020)\citenamefont
  {Girardi-Schappo}, \citenamefont {L}, \citenamefont {Costa}, \citenamefont
  {Carvalho},\ and\ \citenamefont {Kinouchi}}]{Girardi2020}%
  \BibitemOpen
  \bibfield  {author} {\bibinfo {author} {\bibfnamefont {M.}~\bibnamefont
  {Girardi-Schappo}}, \bibinfo {author} {\bibfnamefont {B.}~\bibnamefont {L}},
  \bibinfo {author} {\bibfnamefont {A.~A.}\ \bibnamefont {Costa}}, \bibinfo
  {author} {\bibfnamefont {T.~T.~A.}\ \bibnamefont {Carvalho}}, \ and\ \bibinfo
  {author} {\bibfnamefont {O.}~\bibnamefont {Kinouchi}},\ }\bibfield  {title}
  {\enquote {\bibinfo {title} {Synaptic balance due to homeostatically
  self-organized quasi-critical dynamics},}\ }\href@noop {} {\bibfield
  {journal} {\bibinfo  {journal} {Phys. Rev. Res.}\ }\textbf {\bibinfo {volume}
  {2}},\ \bibinfo {pages} {012042} (\bibinfo {year} {2020})}\BibitemShut
  {NoStop}%
\bibitem [{\citenamefont {Girardi-Schappo}\ \emph {et~al.}(2021)\citenamefont
  {Girardi-Schappo}, \citenamefont {Galera}, \citenamefont {Carvalho},
  \citenamefont {Brochini}, \citenamefont {Kamiji}, \citenamefont {Roque},\
  and\ \citenamefont {Kinouchi}}]{Girardi-Schappo2021}%
  \BibitemOpen
  \bibfield  {author} {\bibinfo {author} {\bibfnamefont {M.}~\bibnamefont
  {Girardi-Schappo}}, \bibinfo {author} {\bibfnamefont {E.~F.}\ \bibnamefont
  {Galera}}, \bibinfo {author} {\bibfnamefont {T.~T.}\ \bibnamefont
  {Carvalho}}, \bibinfo {author} {\bibfnamefont {L.}~\bibnamefont {Brochini}},
  \bibinfo {author} {\bibfnamefont {N.~L.}\ \bibnamefont {Kamiji}}, \bibinfo
  {author} {\bibfnamefont {A.~C.}\ \bibnamefont {Roque}}, \ and\ \bibinfo
  {author} {\bibfnamefont {O.}~\bibnamefont {Kinouchi}},\ }\bibfield  {title}
  {\enquote {\bibinfo {title} {A unified theory of {E/I} synaptic balance,
  quasicritical neuronal avalanches and asynchronous irregular spiking},}\
  }\href {\doibase 10.1088/2632-072X/ac2792} {\bibfield  {journal} {\bibinfo
  {journal} {J. Phys. Complex.}\ }\textbf {\bibinfo {volume} {2}} (\bibinfo
  {year} {2021}),\ 10.1088/2632-072X/ac2792}\BibitemShut {NoStop}%
\bibitem [{\citenamefont {Menesse}\ \emph {et~al.}(2022)\citenamefont
  {Menesse}, \citenamefont {Marin}, \citenamefont {Girardi-Schappo},\ and\
  \citenamefont {Kinouchi}}]{Menesse2021}%
  \BibitemOpen
  \bibfield  {author} {\bibinfo {author} {\bibfnamefont {G.}~\bibnamefont
  {Menesse}}, \bibinfo {author} {\bibfnamefont {B.}~\bibnamefont {Marin}},
  \bibinfo {author} {\bibfnamefont {M.}~\bibnamefont {Girardi-Schappo}}, \ and\
  \bibinfo {author} {\bibfnamefont {O.}~\bibnamefont {Kinouchi}},\ }\bibfield
  {title} {\enquote {\bibinfo {title} {Homeostatic criticality in neuronal
  networks},}\ }\href {\doibase 10.1016/j.chaos.2022.111877} {\bibfield
  {journal} {\bibinfo  {journal} {Chaos Solitons Fractals}\ }\textbf {\bibinfo
  {volume} {156}},\ \bibinfo {pages} {111877} (\bibinfo {year}
  {2022})}\BibitemShut {NoStop}%
\bibitem [{\citenamefont {Kinouchi}, \citenamefont {Pazzini},\ and\
  \citenamefont {Copelli}(2020)}]{Kinouchi2020}%
  \BibitemOpen
  \bibfield  {author} {\bibinfo {author} {\bibfnamefont {O.}~\bibnamefont
  {Kinouchi}}, \bibinfo {author} {\bibfnamefont {R.}~\bibnamefont {Pazzini}}, \
  and\ \bibinfo {author} {\bibfnamefont {M.}~\bibnamefont {Copelli}},\
  }\bibfield  {title} {\enquote {\bibinfo {title} {Mechanisms of self-organized
  quasicriticality in neuronal networks models},}\ }\href@noop {} {\bibfield
  {journal} {\bibinfo  {journal} {Front. Phys.}\ }\textbf {\bibinfo {volume}
  {8}},\ \bibinfo {pages} {583213} (\bibinfo {year} {2020})}\BibitemShut
  {NoStop}%
\bibitem [{\citenamefont {Chialvo}\ \emph {et~al.}(2020)\citenamefont
  {Chialvo}, \citenamefont {Cannas}, \citenamefont {Grigera}, \citenamefont
  {Martin},\ and\ \citenamefont {Plenz}}]{Chialvo2020}%
  \BibitemOpen
  \bibfield  {author} {\bibinfo {author} {\bibfnamefont {D.~R.}\ \bibnamefont
  {Chialvo}}, \bibinfo {author} {\bibfnamefont {S.~A.}\ \bibnamefont {Cannas}},
  \bibinfo {author} {\bibfnamefont {T.~S.}\ \bibnamefont {Grigera}}, \bibinfo
  {author} {\bibfnamefont {D.~A.}\ \bibnamefont {Martin}}, \ and\ \bibinfo
  {author} {\bibfnamefont {D.}~\bibnamefont {Plenz}},\ }\bibfield  {title}
  {\enquote {\bibinfo {title} {Controlling a complex system near its critical
  point via temporal correlations},}\ }\href {\doibase
  10.1038/s41598-020-69154-0} {\bibfield  {journal} {\bibinfo  {journal} {Sci.
  Rep.}\ }\textbf {\bibinfo {volume} {10}},\ \bibinfo {pages} {12145} (\bibinfo
  {year} {2020})}\BibitemShut {NoStop}%
\bibitem [{\citenamefont {Buendia}\ \emph {et~al.}(2020)\citenamefont
  {Buendia}, \citenamefont {di{ }Santo}, \citenamefont {Bonachela},\ and\
  \citenamefont {Muñoz}}]{Buendia2020}%
  \BibitemOpen
  \bibfield  {author} {\bibinfo {author} {\bibfnamefont {V.}~\bibnamefont
  {Buendia}}, \bibinfo {author} {\bibfnamefont {S.}~\bibnamefont {di{ }Santo}},
  \bibinfo {author} {\bibfnamefont {J.~A.}\ \bibnamefont {Bonachela}}, \ and\
  \bibinfo {author} {\bibfnamefont {M.~A.}\ \bibnamefont {Muñoz}},\ }\bibfield
   {title} {\enquote {\bibinfo {title} {Feedback mechanisms for
  self-organization to the edge of a phase transition},}\ }\href {\doibase
  10.3389/fphy.2020.00333} {\bibfield  {journal} {\bibinfo  {journal} {Front.
  Phys.}\ }\textbf {\bibinfo {volume} {8}},\ \bibinfo {pages} {333} (\bibinfo
  {year} {2020})}\BibitemShut {NoStop}%
\bibitem [{\citenamefont {Poil}\ \emph {et~al.}(2012)\citenamefont {Poil},
  \citenamefont {Hardstone}, \citenamefont {Mansvelder},\ and\ \citenamefont
  {Linkenkaer-Hansen}}]{Poil2012}%
  \BibitemOpen
  \bibfield  {author} {\bibinfo {author} {\bibfnamefont {S.-S.}\ \bibnamefont
  {Poil}}, \bibinfo {author} {\bibfnamefont {R.}~\bibnamefont {Hardstone}},
  \bibinfo {author} {\bibfnamefont {H.~D.}\ \bibnamefont {Mansvelder}}, \ and\
  \bibinfo {author} {\bibfnamefont {K.}~\bibnamefont {Linkenkaer-Hansen}},\
  }\bibfield  {title} {\enquote {\bibinfo {title} {Critical-state dynamics of
  avalanches and oscillations jointly emerge from balanced
  excitation/inhibition in neuronal networks},}\ }\href {\doibase
  10.1523/JNEUROSCI.5990-11.2012} {\bibfield  {journal} {\bibinfo  {journal}
  {J. Neurosci.}\ }\textbf {\bibinfo {volume} {32}},\ \bibinfo {pages}
  {9817--9823} (\bibinfo {year} {2012})}\BibitemShut {NoStop}%
\bibitem [{\citenamefont {Di~Santo}\ \emph {et~al.}(2018)\citenamefont
  {Di~Santo}, \citenamefont {Villegas}, \citenamefont {Burioni},\ and\
  \citenamefont {Mu{\~n}oz}}]{DiSanto2018}%
  \BibitemOpen
  \bibfield  {author} {\bibinfo {author} {\bibfnamefont {S.}~\bibnamefont
  {Di~Santo}}, \bibinfo {author} {\bibfnamefont {P.}~\bibnamefont {Villegas}},
  \bibinfo {author} {\bibfnamefont {R.}~\bibnamefont {Burioni}}, \ and\
  \bibinfo {author} {\bibfnamefont {M.~A.}\ \bibnamefont {Mu{\~n}oz}},\
  }\bibfield  {title} {\enquote {\bibinfo {title} {Landau--ginzburg theory of
  cortex dynamics: Scale-free avalanches emerge at the edge of
  synchronization},}\ }\href@noop {} {\bibfield  {journal} {\bibinfo  {journal}
  {Proc Natl Acad Sci USA}\ }\textbf {\bibinfo {volume} {115}},\ \bibinfo
  {pages} {E1356--E1365} (\bibinfo {year} {2018})}\BibitemShut {NoStop}%
\bibitem [{\citenamefont {Dalla{ }Porta}\ and\ \citenamefont
  {Copelli}(2019)}]{DallaPorta2019}%
  \BibitemOpen
  \bibfield  {author} {\bibinfo {author} {\bibfnamefont {L.}~\bibnamefont
  {Dalla{ }Porta}}\ and\ \bibinfo {author} {\bibfnamefont {M.}~\bibnamefont
  {Copelli}},\ }\bibfield  {title} {\enquote {\bibinfo {title} {Modeling
  neuronal avalanches and longrange temporal correlations at the emergence of
  collective oscillations: Continuously varying exponents mimic {M}/{EEG}
  results},}\ }\href {\doibase 10.1371/journal.pcbi.1006924} {\bibfield
  {journal} {\bibinfo  {journal} {PLoS Comput. Biol.}\ }\textbf {\bibinfo
  {volume} {15}},\ \bibinfo {pages} {e1006924} (\bibinfo {year}
  {2019})}\BibitemShut {NoStop}%
\bibitem [{\citenamefont {Buend{\'i}a}\ \emph {et~al.}(2021)\citenamefont
  {Buend{\'i}a}, \citenamefont {Villegas}, \citenamefont {Burioni},\ and\
  \citenamefont {Mu{\~n}oz}}]{Buendia2021}%
  \BibitemOpen
  \bibfield  {author} {\bibinfo {author} {\bibfnamefont {V.}~\bibnamefont
  {Buend{\'i}a}}, \bibinfo {author} {\bibfnamefont {P.}~\bibnamefont
  {Villegas}}, \bibinfo {author} {\bibfnamefont {R.}~\bibnamefont {Burioni}}, \
  and\ \bibinfo {author} {\bibfnamefont {M.~A.}\ \bibnamefont {Mu{\~n}oz}},\
  }\bibfield  {title} {\enquote {\bibinfo {title} {Hybrid-type synchronization
  transitions: Where incipient oscillations, scale-free avalanches, and
  bistability live together},}\ }\href@noop {} {\bibfield  {journal} {\bibinfo
  {journal} {Phys. Rev. Res.}\ }\textbf {\bibinfo {volume} {3}},\ \bibinfo
  {pages} {023224} (\bibinfo {year} {2021})}\BibitemShut {NoStop}%
\bibitem [{\citenamefont {Poil}, \citenamefont {van{ }Ooyen},\ and\
  \citenamefont {Linkenkaer{-}Hansen}(2008)}]{Poil2008}%
  \BibitemOpen
  \bibfield  {author} {\bibinfo {author} {\bibfnamefont {S.-S.}\ \bibnamefont
  {Poil}}, \bibinfo {author} {\bibfnamefont {A.}~\bibnamefont {van{ }Ooyen}}, \
  and\ \bibinfo {author} {\bibfnamefont {K.}~\bibnamefont
  {Linkenkaer{-}Hansen}},\ }\bibfield  {title} {\enquote {\bibinfo {title}
  {Avalanche dynamics of human brain oscillations: relation to critical
  branching processes and temporal correlations},}\ }\href@noop {} {\bibfield
  {journal} {\bibinfo  {journal} {Hum. Brain Mapp.}\ }\textbf {\bibinfo
  {volume} {29}},\ \bibinfo {pages} {770--777} (\bibinfo {year}
  {2008})}\BibitemShut {NoStop}%
\bibitem [{\citenamefont {Kinouchi}\ and\ \citenamefont
  {Tragtenberg}(1996)}]{Kinouchi1996}%
  \BibitemOpen
  \bibfield  {author} {\bibinfo {author} {\bibfnamefont {O.}~\bibnamefont
  {Kinouchi}}\ and\ \bibinfo {author} {\bibfnamefont {M.~H.~R.}\ \bibnamefont
  {Tragtenberg}},\ }\bibfield  {title} {\enquote {\bibinfo {title} {Modeling
  neurons by simple maps},}\ }\href {\doibase 10.1142/S0218127496001508}
  {\bibfield  {journal} {\bibinfo  {journal} {Int J Bifurcat Chaos}\ }\textbf
  {\bibinfo {volume} {6}},\ \bibinfo {pages} {2343--2360} (\bibinfo {year}
  {1996})}\BibitemShut {NoStop}%
\bibitem [{\citenamefont {Girardi-Schappo}\ \emph {et~al.}(2017)\citenamefont
  {Girardi-Schappo}, \citenamefont {Bortolotto}, \citenamefont {Stenzinger},
  \citenamefont {Gonsalves},\ and\ \citenamefont {Tragtenberg}}]{Girardi2017}%
  \BibitemOpen
  \bibfield  {author} {\bibinfo {author} {\bibfnamefont {M.}~\bibnamefont
  {Girardi-Schappo}}, \bibinfo {author} {\bibfnamefont {G.~S.}\ \bibnamefont
  {Bortolotto}}, \bibinfo {author} {\bibfnamefont {R.~V.}\ \bibnamefont
  {Stenzinger}}, \bibinfo {author} {\bibfnamefont {J.~J.}\ \bibnamefont
  {Gonsalves}}, \ and\ \bibinfo {author} {\bibfnamefont {M.~H.}\ \bibnamefont
  {Tragtenberg}},\ }\bibfield  {title} {\enquote {\bibinfo {title} {Phase
  diagrams and dynamics of a computationally efficient map-based neuron
  model},}\ }\href@noop {} {\bibfield  {journal} {\bibinfo  {journal} {{PLoS}
  One}\ }\textbf {\bibinfo {volume} {12}},\ \bibinfo {pages} {e0174621}
  (\bibinfo {year} {2017})}\BibitemShut {NoStop}%
\bibitem [{\citenamefont {Courbage}\ and\ \citenamefont
  {Nekorkin}(2010)}]{Courbage2010}%
  \BibitemOpen
  \bibfield  {author} {\bibinfo {author} {\bibfnamefont {M.}~\bibnamefont
  {Courbage}}\ and\ \bibinfo {author} {\bibfnamefont {V.~I.}\ \bibnamefont
  {Nekorkin}},\ }\bibfield  {title} {\enquote {\bibinfo {title} {Map-based
  models in neurodynamics},}\ }\href@noop {} {\bibfield  {journal} {\bibinfo
  {journal} {Int J Bifurcat Chaos}\ }\textbf {\bibinfo {volume} {20}},\
  \bibinfo {pages} {1631--1651} (\bibinfo {year} {2010})}\BibitemShut {NoStop}%
\bibitem [{\citenamefont {Ibarz}, \citenamefont {Casado},\ and\ \citenamefont
  {Sanju{\'a}n}(2011)}]{Ibarz2011}%
  \BibitemOpen
  \bibfield  {author} {\bibinfo {author} {\bibfnamefont {B.}~\bibnamefont
  {Ibarz}}, \bibinfo {author} {\bibfnamefont {J.~M.}\ \bibnamefont {Casado}}, \
  and\ \bibinfo {author} {\bibfnamefont {M.~A.~F.}\ \bibnamefont
  {Sanju{\'a}n}},\ }\bibfield  {title} {\enquote {\bibinfo {title} {Map-based
  models in neuronal dynamics},}\ }\href@noop {} {\bibfield  {journal}
  {\bibinfo  {journal} {Phys. Rep.}\ }\textbf {\bibinfo {volume} {501}},\
  \bibinfo {pages} {1--74} (\bibinfo {year} {2011})}\BibitemShut {NoStop}%
\bibitem [{\citenamefont {Girardi{-}Schappo}, \citenamefont {Tragtenberg},\
  and\ \citenamefont {Kinouchi}(2013)}]{Girardi2013}%
  \BibitemOpen
  \bibfield  {author} {\bibinfo {author} {\bibfnamefont {M.}~\bibnamefont
  {Girardi{-}Schappo}}, \bibinfo {author} {\bibfnamefont {M.}~\bibnamefont
  {Tragtenberg}}, \ and\ \bibinfo {author} {\bibfnamefont {O.}~\bibnamefont
  {Kinouchi}},\ }\bibfield  {title} {\enquote {\bibinfo {title} {A brief
  history of excitable map-based neurons and neural networks},}\ }\href@noop {}
  {\bibfield  {journal} {\bibinfo  {journal} {J. Neurosci. Methods}\ }\textbf
  {\bibinfo {volume} {220}},\ \bibinfo {pages} {116--130} (\bibinfo {year}
  {2013})}\BibitemShut {NoStop}%
\bibitem [{\citenamefont {Kuva}\ \emph {et~al.}(2001)\citenamefont {Kuva},
  \citenamefont {Lima}, \citenamefont {Kinouchi}, \citenamefont {Tragtenberg},\
  and\ \citenamefont {Roque}}]{Kuva2001}%
  \BibitemOpen
  \bibfield  {author} {\bibinfo {author} {\bibfnamefont {S.~M.}\ \bibnamefont
  {Kuva}}, \bibinfo {author} {\bibfnamefont {G.~F.}\ \bibnamefont {Lima}},
  \bibinfo {author} {\bibfnamefont {O.}~\bibnamefont {Kinouchi}}, \bibinfo
  {author} {\bibfnamefont {M.~H.}\ \bibnamefont {Tragtenberg}}, \ and\ \bibinfo
  {author} {\bibfnamefont {A.~C.}\ \bibnamefont {Roque}},\ }\bibfield  {title}
  {\enquote {\bibinfo {title} {A minimal model for excitable and bursting
  elements},}\ }\href {\doibase 10.1016/S0925-2312(01)00376-9} {\bibfield
  {journal} {\bibinfo  {journal} {Neurocomputing}\ }\textbf {\bibinfo {volume}
  {38-40}},\ \bibinfo {pages} {255--261} (\bibinfo {year} {2001})}\BibitemShut
  {NoStop}%
\bibitem [{\citenamefont {Copelli}, \citenamefont {Tragtenberg},\ and\
  \citenamefont {Kinouchi}(2004)}]{Copelli2004}%
  \BibitemOpen
  \bibfield  {author} {\bibinfo {author} {\bibfnamefont {M.}~\bibnamefont
  {Copelli}}, \bibinfo {author} {\bibfnamefont {M.}~\bibnamefont
  {Tragtenberg}}, \ and\ \bibinfo {author} {\bibfnamefont {O.}~\bibnamefont
  {Kinouchi}},\ }\bibfield  {title} {\enquote {\bibinfo {title} {Stability
  diagrams for bursting neurons modeled by three-variable maps},}\ }\href
  {\doibase 10.1016/j.physa.2004.04.087} {\bibfield  {journal} {\bibinfo
  {journal} {Physica A}\ }\textbf {\bibinfo {volume} {342}},\ \bibinfo {pages}
  {263--269} (\bibinfo {year} {2004})}\BibitemShut {NoStop}%
\bibitem [{\citenamefont {Girardi{-}Schappo}, \citenamefont {Kinouchi},\ and\
  \citenamefont {Tragtenberg}(2013)}]{Girardi2013b}%
  \BibitemOpen
  \bibfield  {author} {\bibinfo {author} {\bibfnamefont {M.}~\bibnamefont
  {Girardi{-}Schappo}}, \bibinfo {author} {\bibfnamefont {O.}~\bibnamefont
  {Kinouchi}}, \ and\ \bibinfo {author} {\bibfnamefont {M.~H.~R.}\ \bibnamefont
  {Tragtenberg}},\ }\bibfield  {title} {\enquote {\bibinfo {title} {Critical
  avalanches and subsampling in map-based neural networks coupled with noisy
  synapses},}\ }\href@noop {} {\bibfield  {journal} {\bibinfo  {journal} {Phys.
  Rev. E}\ }\textbf {\bibinfo {volume} {88}},\ \bibinfo {pages} {024701}
  (\bibinfo {year} {2013})}\BibitemShut {NoStop}%
\bibitem [{\citenamefont {Yokoi}, \citenamefont {de{ }Oliveira},\ and\
  \citenamefont {Salinas}(1985)}]{Yokoi1985}%
  \BibitemOpen
  \bibfield  {author} {\bibinfo {author} {\bibfnamefont {C.~S.~O.}\
  \bibnamefont {Yokoi}}, \bibinfo {author} {\bibfnamefont {M.~J.}\ \bibnamefont
  {de{ }Oliveira}}, \ and\ \bibinfo {author} {\bibfnamefont {S.~R.}\
  \bibnamefont {Salinas}},\ }\bibfield  {title} {\enquote {\bibinfo {title}
  {Strange attractor in the {I}sing model with competing interactions on the
  {C}ayley tree},}\ }\href@noop {} {\bibfield  {journal} {\bibinfo  {journal}
  {Phys. Rev. Lett.}\ }\textbf {\bibinfo {volume} {54(3)}},\ \bibinfo {pages}
  {163--166} (\bibinfo {year} {1985})}\BibitemShut {NoStop}%
\bibitem [{\citenamefont {Tragtenberg}\ and\ \citenamefont
  {Yokoi}(1995)}]{Tragtenberg1995}%
  \BibitemOpen
  \bibfield  {author} {\bibinfo {author} {\bibfnamefont {M.~H.~R.}\
  \bibnamefont {Tragtenberg}}\ and\ \bibinfo {author} {\bibfnamefont
  {C.~S.~O.}\ \bibnamefont {Yokoi}},\ }\bibfield  {title} {\enquote {\bibinfo
  {title} {Field behavior of an {I}sing model with competing interactions on
  the {B}ethe lattice},}\ }\href@noop {} {\bibfield  {journal} {\bibinfo
  {journal} {Phys. Rev. E}\ }\textbf {\bibinfo {volume} {52(3)}},\ \bibinfo
  {pages} {2187--2197} (\bibinfo {year} {1995})}\BibitemShut {NoStop}%
\bibitem [{\citenamefont {Kinouchi}(2001)}]{kinouchi2001map}%
  \BibitemOpen
  \bibfield  {author} {\bibinfo {author} {\bibfnamefont {O.}~\bibnamefont
  {Kinouchi}},\ }\bibfield  {title} {\enquote {\bibinfo {title} {Extended
  dynamical range as a collective property of excitable cells},}\ }\href
  {\doibase 10.48550/arXiv.cond-mat/0108404} {\bibfield  {journal} {\bibinfo
  {journal} {arXiv:cond-mat/0108404 [cond-mat.dis-nn]}\ } (\bibinfo {year}
  {2001}),\ 10.48550/arXiv.cond-mat/0108404}\BibitemShut {NoStop}%
\bibitem [{\citenamefont {Hindmarsh}\ and\ \citenamefont
  {Rose}(1984)}]{Hindmarsh1984}%
  \BibitemOpen
  \bibfield  {author} {\bibinfo {author} {\bibfnamefont {J.~L.}\ \bibnamefont
  {Hindmarsh}}\ and\ \bibinfo {author} {\bibfnamefont {R.}~\bibnamefont
  {Rose}},\ }\bibfield  {title} {\enquote {\bibinfo {title} {A model of
  neuronal bursting using three coupled first order differential equations},}\
  }\href@noop {} {\bibfield  {journal} {\bibinfo  {journal} {Proc R Soc Lond B
  Biol Sci .}\ }\textbf {\bibinfo {volume} {221}},\ \bibinfo {pages} {87--102}
  (\bibinfo {year} {1984})}\BibitemShut {NoStop}%
\bibitem [{\citenamefont {Strogatz}(2001)}]{strogatzBook}%
  \BibitemOpen
  \bibfield  {author} {\bibinfo {author} {\bibfnamefont {S.~H.}\ \bibnamefont
  {Strogatz}},\ }\href@noop {} {\emph {\bibinfo {title} {Nonlinear Dynamics and
  Chaos: With Applications to Physics, Biology, Chemistry, and Engineering}}}\
  (\bibinfo  {publisher} {Westview Press},\ \bibinfo {year} {2001})\BibitemShut
  {NoStop}%
\bibitem [{\citenamefont {Roth}\ and\ \citenamefont
  {van{~}Rossum}(2010)}]{RothRossum2010}%
  \BibitemOpen
  \bibfield  {author} {\bibinfo {author} {\bibfnamefont {A.}~\bibnamefont
  {Roth}}\ and\ \bibinfo {author} {\bibfnamefont {M.~C.~W.}\ \bibnamefont
  {van{~}Rossum}},\ }\bibfield  {title} {\enquote {\bibinfo {title} {Modeling
  synapses},}\ }in\ \href@noop {} {\emph {\bibinfo {booktitle} {Computational
  Modeling Methods for Neurocientists}}},\ \bibinfo {editor} {edited by\
  \bibinfo {editor} {\bibfnamefont {E.}~\bibnamefont {de{~}Schutter}}}\
  (\bibinfo  {publisher} {The MIT Press},\ \bibinfo {address} {Cambridge, MA,
  USA},\ \bibinfo {year} {2010})\BibitemShut {NoStop}%
\bibitem [{\citenamefont {Kuramoto}(1984)}]{Kuramoto1984}%
  \BibitemOpen
  \bibfield  {author} {\bibinfo {author} {\bibfnamefont {Y.}~\bibnamefont
  {Kuramoto}},\ }\href {\doibase 10.1007/978-3-642-69689-3} {\emph {\bibinfo
  {title} {Chemical Oscillations, Waves, and Turbulence}}},\ Vol.~\bibinfo
  {volume} {19}\ (\bibinfo  {publisher} {Springer Berlin Heidelberg},\ \bibinfo
  {year} {1984})\BibitemShut {NoStop}%
\bibitem [{\citenamefont {Brunel}(2000)}]{Brunel2000}%
  \BibitemOpen
  \bibfield  {author} {\bibinfo {author} {\bibfnamefont {N.}~\bibnamefont
  {Brunel}},\ }\bibfield  {title} {\enquote {\bibinfo {title} {Dynamics of
  sparsely connected networks of excitatory and inhibitory spiking neurons},}\
  }\href@noop {} {\bibfield  {journal} {\bibinfo  {journal} {J. Comput.
  Neurosci.}\ }\textbf {\bibinfo {volume} {8}},\ \bibinfo {pages} {183--208}
  (\bibinfo {year} {2000})}\BibitemShut {NoStop}%
\bibitem [{\citenamefont {Golomb}\ and\ \citenamefont
  {Rinzel}(1994)}]{Golomb1994}%
  \BibitemOpen
  \bibfield  {author} {\bibinfo {author} {\bibfnamefont {D.}~\bibnamefont
  {Golomb}}\ and\ \bibinfo {author} {\bibfnamefont {J.}~\bibnamefont
  {Rinzel}},\ }\bibfield  {title} {\enquote {\bibinfo {title} {Clustering in
  globally coupled inhibitory neurons},}\ }\href@noop {} {\bibfield  {journal}
  {\bibinfo  {journal} {Physica D}\ }\textbf {\bibinfo {volume} {72}},\
  \bibinfo {pages} {259--282} (\bibinfo {year} {1994})}\BibitemShut {NoStop}%
\bibitem [{\citenamefont {Golomb}(2007)}]{Golomb2007}%
  \BibitemOpen
  \bibfield  {author} {\bibinfo {author} {\bibfnamefont {D.}~\bibnamefont
  {Golomb}},\ }\bibfield  {title} {\enquote {\bibinfo {title} {{N}euronal
  synchrony measures},}\ }\href {\doibase 10.4249/scholarpedia.1347} {\bibfield
   {journal} {\bibinfo  {journal} {Scholarpedia}\ }\textbf {\bibinfo {volume}
  {2}},\ \bibinfo {pages} {1347} (\bibinfo {year} {2007})},\ \bibinfo {note}
  {revision \#128277}\BibitemShut {NoStop}%
\bibitem [{\citenamefont {Brochini}\ \emph {et~al.}(2016)\citenamefont
  {Brochini}, \citenamefont {Costa}, \citenamefont {Abadi}, \citenamefont
  {Roque}, \citenamefont {Stolfi},\ and\ \citenamefont
  {Kinouchi}}]{Brochini2016}%
  \BibitemOpen
  \bibfield  {author} {\bibinfo {author} {\bibfnamefont {L.}~\bibnamefont
  {Brochini}}, \bibinfo {author} {\bibfnamefont {A.~A.}\ \bibnamefont {Costa}},
  \bibinfo {author} {\bibfnamefont {M.}~\bibnamefont {Abadi}}, \bibinfo
  {author} {\bibfnamefont {A.~C.}\ \bibnamefont {Roque}}, \bibinfo {author}
  {\bibfnamefont {J.}~\bibnamefont {Stolfi}}, \ and\ \bibinfo {author}
  {\bibfnamefont {O.}~\bibnamefont {Kinouchi}},\ }\bibfield  {title} {\enquote
  {\bibinfo {title} {Phase transitions and self-organized criticality in
  networks of stochastic spiking neurons},}\ }\href {\doibase
  10.1038/srep35831} {\bibfield  {journal} {\bibinfo  {journal} {Sci. Rep.}\
  }\textbf {\bibinfo {volume} {6}},\ \bibinfo {pages} {35831} (\bibinfo {year}
  {2016})}\BibitemShut {NoStop}%
\bibitem [{\citenamefont {Costa}\ \emph {et~al.}(2018)\citenamefont {Costa},
  \citenamefont {Amon}, \citenamefont {Sporns},\ and\ \citenamefont
  {Favela}}]{Costa2018}%
  \BibitemOpen
  \bibfield  {author} {\bibinfo {author} {\bibfnamefont {A.~A.}\ \bibnamefont
  {Costa}}, \bibinfo {author} {\bibfnamefont {M.~J.}\ \bibnamefont {Amon}},
  \bibinfo {author} {\bibfnamefont {O.}~\bibnamefont {Sporns}}, \ and\ \bibinfo
  {author} {\bibfnamefont {L.~H.}\ \bibnamefont {Favela}},\ }\bibfield  {title}
  {\enquote {\bibinfo {title} {Fractal analyses of networks of
  integrate-and-fire stochastic spiking neurons},}\ }in\ \href@noop {} {\emph
  {\bibinfo {booktitle} {International Workshop on Complex Networks}}}\
  (\bibinfo {organization} {Springer},\ \bibinfo {year} {2018})\ pp.\ \bibinfo
  {pages} {161--171}\BibitemShut {NoStop}%
\bibitem [{\citenamefont {Menesse}\ and\ \citenamefont
  {Kinouchi}(2023)}]{Menesse2023}%
  \BibitemOpen
  \bibfield  {author} {\bibinfo {author} {\bibfnamefont {G.}~\bibnamefont
  {Menesse}}\ and\ \bibinfo {author} {\bibfnamefont {O.}~\bibnamefont
  {Kinouchi}},\ }\bibfield  {title} {\enquote {\bibinfo {title} {Less is
  different: Why sparse networks with inhibition differ from complete
  graphs},}\ }\href {\doibase 10.1103/PhysRevE.108.024315} {\bibfield
  {journal} {\bibinfo  {journal} {Phys. Rev. E}\ }\textbf {\bibinfo {volume}
  {108}},\ \bibinfo {pages} {024315} (\bibinfo {year} {2023})}\BibitemShut
  {NoStop}%
\bibitem [{\citenamefont {Holthoff}, \citenamefont {Kovalchuk},\ and\
  \citenamefont {Konnerth}(2006)}]{Holthoff2006}%
  \BibitemOpen
  \bibfield  {author} {\bibinfo {author} {\bibfnamefont {K.}~\bibnamefont
  {Holthoff}}, \bibinfo {author} {\bibfnamefont {Y.}~\bibnamefont {Kovalchuk}},
  \ and\ \bibinfo {author} {\bibfnamefont {A.}~\bibnamefont {Konnerth}},\
  }\bibfield  {title} {\enquote {\bibinfo {title} {Dendritic spikes and
  activity-dependent synaptic plasticity},}\ }\href@noop {} {\bibfield
  {journal} {\bibinfo  {journal} {Cell Tissue Res}\ }\textbf {\bibinfo {volume}
  {326}},\ \bibinfo {pages} {369--377} (\bibinfo {year} {2006})}\BibitemShut
  {NoStop}%
\bibitem [{\citenamefont {Gollo}, \citenamefont {Kinouchi},\ and\ \citenamefont
  {Copelli}(2009)}]{Gollo2009}%
  \BibitemOpen
  \bibfield  {author} {\bibinfo {author} {\bibfnamefont {L.~L.}\ \bibnamefont
  {Gollo}}, \bibinfo {author} {\bibfnamefont {O.}~\bibnamefont {Kinouchi}}, \
  and\ \bibinfo {author} {\bibfnamefont {M.}~\bibnamefont {Copelli}},\
  }\bibfield  {title} {\enquote {\bibinfo {title} {Active dendrites enhance
  neuronal dynamic range},}\ }\href@noop {} {\bibfield  {journal} {\bibinfo
  {journal} {PLos Comput. Biol.}\ }\textbf {\bibinfo {volume} {5}},\ \bibinfo
  {pages} {e1000402} (\bibinfo {year} {2009})}\BibitemShut {NoStop}%
\bibitem [{\citenamefont {Shimoura}\ \emph {et~al.}(2021)\citenamefont
  {Shimoura}, \citenamefont {Pena}, \citenamefont {Lima}, \citenamefont
  {Kamiji}, \citenamefont {Girardi-Schappo},\ and\ \citenamefont
  {Roque}}]{Shimoura2021}%
  \BibitemOpen
  \bibfield  {author} {\bibinfo {author} {\bibfnamefont {R.~O.}\ \bibnamefont
  {Shimoura}}, \bibinfo {author} {\bibfnamefont {R.~F.~O.}\ \bibnamefont
  {Pena}}, \bibinfo {author} {\bibfnamefont {V.}~\bibnamefont {Lima}}, \bibinfo
  {author} {\bibfnamefont {N.~L.}\ \bibnamefont {Kamiji}}, \bibinfo {author}
  {\bibfnamefont {M.}~\bibnamefont {Girardi-Schappo}}, \ and\ \bibinfo {author}
  {\bibfnamefont {A.~C.}\ \bibnamefont {Roque}},\ }\bibfield  {title} {\enquote
  {\bibinfo {title} {Building a model of the brain: from detailed connectivity
  maps to network organization},}\ }\href {\doibase
  10.1140/epjs/s11734-021-00152-7} {\bibfield  {journal} {\bibinfo  {journal}
  {Eur Phys J Spec Top}\ }\textbf {\bibinfo {volume} {230}},\ \bibinfo {pages}
  {2887--2909} (\bibinfo {year} {2021})}\BibitemShut {NoStop}%
\bibitem [{\citenamefont {Graf}\ and\ \citenamefont
  {Machta}(2024)}]{Graf2024Bif}%
  \BibitemOpen
  \bibfield  {author} {\bibinfo {author} {\bibfnamefont {I.~R.}\ \bibnamefont
  {Graf}}\ and\ \bibinfo {author} {\bibfnamefont {B.~B.}\ \bibnamefont
  {Machta}},\ }\bibfield  {title} {\enquote {\bibinfo {title} {A bifurcation
  integrates information from many noisy ion channels and allows for
  milli-kelvin thermal sensitivity in the snake pit organ},}\ }\href {\doibase
  10.1073/pnas.2308215121} {\bibfield  {journal} {\bibinfo  {journal} {Proc.
  Natl. Acad. Sci. USA}\ }\textbf {\bibinfo {volume} {121}},\ \bibinfo {pages}
  {e2308215121} (\bibinfo {year} {2024})}\BibitemShut {NoStop}%
\bibitem [{\citenamefont {Bonachela}\ \emph {et~al.}(2010)\citenamefont
  {Bonachela}, \citenamefont {de{ }Franciscis}, \citenamefont {Torres},\ and\
  \citenamefont {Mu{\~n}oz}}]{Bonachela2010}%
  \BibitemOpen
  \bibfield  {author} {\bibinfo {author} {\bibfnamefont {J.~A.}\ \bibnamefont
  {Bonachela}}, \bibinfo {author} {\bibfnamefont {S.}~\bibnamefont {de{
  }Franciscis}}, \bibinfo {author} {\bibfnamefont {J.~J.}\ \bibnamefont
  {Torres}}, \ and\ \bibinfo {author} {\bibfnamefont {M.~A.}\ \bibnamefont
  {Mu{\~n}oz}},\ }\bibfield  {title} {\enquote {\bibinfo {title}
  {Self-organization without conservation: are neuronal avalanches generically
  critical?}}\ }\href@noop {} {\bibfield  {journal} {\bibinfo  {journal} {J.
  Stat. Mech.}\ }\textbf {\bibinfo {volume} {2010}},\ \bibinfo {pages} {P02015}
  (\bibinfo {year} {2010})}\BibitemShut {NoStop}%
\bibitem [{\citenamefont {van Ede}\ \emph {et~al.}(2018)\citenamefont {van
  Ede}, \citenamefont {Quinn}, \citenamefont {Woolrich},\ and\ \citenamefont
  {Nobre}}]{van2018}%
  \BibitemOpen
  \bibfield  {author} {\bibinfo {author} {\bibfnamefont {F.}~\bibnamefont {van
  Ede}}, \bibinfo {author} {\bibfnamefont {A.~J.}\ \bibnamefont {Quinn}},
  \bibinfo {author} {\bibfnamefont {M.~W.}\ \bibnamefont {Woolrich}}, \ and\
  \bibinfo {author} {\bibfnamefont {A.~C.}\ \bibnamefont {Nobre}},\ }\bibfield
  {title} {\enquote {\bibinfo {title} {Neural oscillations: Sustained rhythms
  or transient burst-events?}}\ }\href {\doibase 10.1016/j.tins.2018.04.004}
  {\bibfield  {journal} {\bibinfo  {journal} {Trends Neurosci.}\ }\textbf
  {\bibinfo {volume} {41}},\ \bibinfo {pages} {415--417} (\bibinfo {year}
  {2018})}\BibitemShut {NoStop}%
\bibitem [{\citenamefont {Tal}\ \emph {et~al.}(2020)\citenamefont {Tal},
  \citenamefont {Neymotin}, \citenamefont {Bickel}, \citenamefont {Lakatos},\
  and\ \citenamefont {Schroeder}}]{Tal2020}%
  \BibitemOpen
  \bibfield  {author} {\bibinfo {author} {\bibfnamefont {I.}~\bibnamefont
  {Tal}}, \bibinfo {author} {\bibfnamefont {S.}~\bibnamefont {Neymotin}},
  \bibinfo {author} {\bibfnamefont {S.}~\bibnamefont {Bickel}}, \bibinfo
  {author} {\bibfnamefont {P.}~\bibnamefont {Lakatos}}, \ and\ \bibinfo
  {author} {\bibfnamefont {C.~E.}\ \bibnamefont {Schroeder}},\ }\bibfield
  {title} {\enquote {\bibinfo {title} {Oscillatory bursting as a mechanism for
  temporal coupling and information coding},}\ }\href {\doibase
  10.3389/fncom.2020.00082} {\bibfield  {journal} {\bibinfo  {journal} {Front
  Comput Neurosci}\ }\textbf {\bibinfo {volume} {14}},\ \bibinfo {pages} {82}
  (\bibinfo {year} {2020})}\BibitemShut {NoStop}%
\bibitem [{\citenamefont {Schmidt}, \citenamefont {Rose},\ and\ \citenamefont
  {Muralidharan}(2023)}]{Schmidt2023}%
  \BibitemOpen
  \bibfield  {author} {\bibinfo {author} {\bibfnamefont {R.}~\bibnamefont
  {Schmidt}}, \bibinfo {author} {\bibfnamefont {J.}~\bibnamefont {Rose}}, \
  and\ \bibinfo {author} {\bibfnamefont {V.}~\bibnamefont {Muralidharan}},\
  }\bibfield  {title} {\enquote {\bibinfo {title} {Transient oscillations as
  computations for cognition: Analysis, modeling and function},}\ }\href
  {\doibase 10.1016/j.conb.2023.102796} {\bibfield  {journal} {\bibinfo
  {journal} {Curr Opin Neurobiol.}\ }\textbf {\bibinfo {volume} {83}},\
  \bibinfo {pages} {102796} (\bibinfo {year} {2023})}\BibitemShut {NoStop}%
\bibitem [{\citenamefont {Costa}, \citenamefont {Brochini},\ and\ \citenamefont
  {Kinouchi}(2017)}]{Costa2017}%
  \BibitemOpen
  \bibfield  {author} {\bibinfo {author} {\bibfnamefont {A.~A.}\ \bibnamefont
  {Costa}}, \bibinfo {author} {\bibfnamefont {L.}~\bibnamefont {Brochini}}, \
  and\ \bibinfo {author} {\bibfnamefont {O.}~\bibnamefont {Kinouchi}},\
  }\bibfield  {title} {\enquote {\bibinfo {title} {Self-organized
  supercriticality and oscillations in networks of stochastic spiking
  neurons},}\ }\href {\doibase 10.3390/e19080399} {\bibfield  {journal}
  {\bibinfo  {journal} {Entropy}\ }\textbf {\bibinfo {volume} {19}},\ \bibinfo
  {pages} {399} (\bibinfo {year} {2017})}\BibitemShut {NoStop}%
\bibitem [{\citenamefont {Kaneko}(1994)}]{kaneko1994CML}%
  \BibitemOpen
  \bibfield  {author} {\bibinfo {author} {\bibfnamefont {K.}~\bibnamefont
  {Kaneko}},\ }\bibfield  {title} {\enquote {\bibinfo {title} {Relevance of
  dynamic clustering to biological networks},}\ }\href@noop {} {\bibfield
  {journal} {\bibinfo  {journal} {Physica D}\ }\textbf {\bibinfo {volume}
  {75}},\ \bibinfo {pages} {55--73} (\bibinfo {year} {1994})}\BibitemShut
  {NoStop}%
\bibitem [{\citenamefont {Kaneko}(1993)}]{kaneko1993CMLbook}%
  \BibitemOpen
  \bibfield  {author} {\bibinfo {author} {\bibfnamefont {K.}~\bibnamefont
  {Kaneko}},\ }\href@noop {} {\emph {\bibinfo {title} {Theory and Applications
  of Coupled Map Lattices}}}\ (\bibinfo  {publisher} {Wiley},\ \bibinfo {year}
  {1993})\BibitemShut {NoStop}%
\bibitem [{\citenamefont {Hernandez{-U}rbina}\ and\ \citenamefont
  {Herrmann}(2017)}]{hernandez2017self}%
  \BibitemOpen
  \bibfield  {author} {\bibinfo {author} {\bibfnamefont {V.}~\bibnamefont
  {Hernandez{-U}rbina}}\ and\ \bibinfo {author} {\bibfnamefont {J.~M.}\
  \bibnamefont {Herrmann}},\ }\bibfield  {title} {\enquote {\bibinfo {title}
  {Self-organized criticality via retro-synaptic signals},}\ }\href@noop {}
  {\bibfield  {journal} {\bibinfo  {journal} {Front. Phys.}\ }\textbf {\bibinfo
  {volume} {4}},\ \bibinfo {pages} {54} (\bibinfo {year} {2017})}\BibitemShut
  {NoStop}%
\bibitem [{\citenamefont {Dickman}\ \emph {et~al.}(2000)\citenamefont
  {Dickman}, \citenamefont {Mu{\~n}oz}, \citenamefont {Vespignani},\ and\
  \citenamefont {Zapperi}}]{Dickman2000}%
  \BibitemOpen
  \bibfield  {author} {\bibinfo {author} {\bibfnamefont {R.}~\bibnamefont
  {Dickman}}, \bibinfo {author} {\bibfnamefont {M.~A.}\ \bibnamefont
  {Mu{\~n}oz}}, \bibinfo {author} {\bibfnamefont {A.}~\bibnamefont
  {Vespignani}}, \ and\ \bibinfo {author} {\bibfnamefont {S.}~\bibnamefont
  {Zapperi}},\ }\bibfield  {title} {\enquote {\bibinfo {title} {Paths to
  self-organized criticality},}\ }\href@noop {} {\bibfield  {journal} {\bibinfo
   {journal} {Braz. J. Phys.}\ }\textbf {\bibinfo {volume} {30}},\ \bibinfo
  {pages} {27--41} (\bibinfo {year} {2000})}\BibitemShut {NoStop}%
\bibitem [{\citenamefont {Bak}, \citenamefont {Tang},\ and\ \citenamefont
  {Wiesenfeld}(1987)}]{Bak1987}%
  \BibitemOpen
  \bibfield  {author} {\bibinfo {author} {\bibfnamefont {P.}~\bibnamefont
  {Bak}}, \bibinfo {author} {\bibfnamefont {C.}~\bibnamefont {Tang}}, \ and\
  \bibinfo {author} {\bibfnamefont {K.}~\bibnamefont {Wiesenfeld}},\ }\bibfield
   {title} {\enquote {\bibinfo {title} {Self-organized criticality: An
  explanation of the 1/f noise},}\ }\href@noop {} {\bibfield  {journal}
  {\bibinfo  {journal} {Phys. Rev. Lett.}\ }\textbf {\bibinfo {volume} {59}},\
  \bibinfo {pages} {381} (\bibinfo {year} {1987})}\BibitemShut {NoStop}%
\bibitem [{\citenamefont {Jiruska}\ \emph {et~al.}(2013)\citenamefont
  {Jiruska}, \citenamefont {De~Curtis}, \citenamefont {Jefferys}, \citenamefont
  {Schevon}, \citenamefont {Schiff},\ and\ \citenamefont
  {Schindler}}]{jiruska2013synchronization}%
  \BibitemOpen
  \bibfield  {author} {\bibinfo {author} {\bibfnamefont {P.}~\bibnamefont
  {Jiruska}}, \bibinfo {author} {\bibfnamefont {M.}~\bibnamefont {De~Curtis}},
  \bibinfo {author} {\bibfnamefont {J.~G.}\ \bibnamefont {Jefferys}}, \bibinfo
  {author} {\bibfnamefont {C.~A.}\ \bibnamefont {Schevon}}, \bibinfo {author}
  {\bibfnamefont {S.~J.}\ \bibnamefont {Schiff}}, \ and\ \bibinfo {author}
  {\bibfnamefont {K.}~\bibnamefont {Schindler}},\ }\bibfield  {title} {\enquote
  {\bibinfo {title} {Synchronization and desynchronization in epilepsy:
  controversies and hypotheses},}\ }\href@noop {} {\bibfield  {journal}
  {\bibinfo  {journal} {J Physiol.}\ }\textbf {\bibinfo {volume} {591}},\
  \bibinfo {pages} {787--797} (\bibinfo {year} {2013})}\BibitemShut {NoStop}%
\bibitem [{\citenamefont {Bernhardt}\ \emph {et~al.}(2013)\citenamefont
  {Bernhardt}, \citenamefont {Hong}, \citenamefont {Bernasconi},\ and\
  \citenamefont {Bernasconi}}]{Bernhardt2013}%
  \BibitemOpen
  \bibfield  {author} {\bibinfo {author} {\bibfnamefont {B.~C.}\ \bibnamefont
  {Bernhardt}}, \bibinfo {author} {\bibfnamefont {S.}~\bibnamefont {Hong}},
  \bibinfo {author} {\bibfnamefont {A.}~\bibnamefont {Bernasconi}}, \ and\
  \bibinfo {author} {\bibfnamefont {N.}~\bibnamefont {Bernasconi}},\ }\bibfield
   {title} {\enquote {\bibinfo {title} {Imaging structural and functional brain
  networks in temporal lobe epilepsy},}\ }\href {\doibase
  10.3389/fnhum.2013.00624} {\bibfield  {journal} {\bibinfo  {journal} {Front
  Hum Neurosci.}\ }\textbf {\bibinfo {volume} {7}},\ \bibinfo {pages} {624}
  (\bibinfo {year} {2013})}\BibitemShut {NoStop}%
\bibitem [{\citenamefont {Girardi{-}Schappo}\ \emph {et~al.}(2021)\citenamefont
  {Girardi{-}Schappo}, \citenamefont {Fadaie}, \citenamefont {Lee},
  \citenamefont {Caldairou}, \citenamefont {Sziklas}, \citenamefont {Crane},
  \citenamefont {Bernhardt}, \citenamefont {Bernasconi},\ and\ \citenamefont
  {Bernasconi}}]{Girardi2021Epil}%
  \BibitemOpen
  \bibfield  {author} {\bibinfo {author} {\bibfnamefont {M.}~\bibnamefont
  {Girardi{-}Schappo}}, \bibinfo {author} {\bibfnamefont {F.}~\bibnamefont
  {Fadaie}}, \bibinfo {author} {\bibfnamefont {H.~M.}\ \bibnamefont {Lee}},
  \bibinfo {author} {\bibfnamefont {B.}~\bibnamefont {Caldairou}}, \bibinfo
  {author} {\bibfnamefont {V.}~\bibnamefont {Sziklas}}, \bibinfo {author}
  {\bibfnamefont {J.}~\bibnamefont {Crane}}, \bibinfo {author} {\bibfnamefont
  {B.~C.}\ \bibnamefont {Bernhardt}}, \bibinfo {author} {\bibfnamefont
  {A.}~\bibnamefont {Bernasconi}}, \ and\ \bibinfo {author} {\bibfnamefont
  {N.}~\bibnamefont {Bernasconi}},\ }\bibfield  {title} {\enquote {\bibinfo
  {title} {Altered communication dynamics reflect cognitive deficits in
  temporal lobe epilepsy},}\ }\href {\doibase 10.1111/epi.16864} {\bibfield
  {journal} {\bibinfo  {journal} {Epilepsia}\ }\textbf {\bibinfo {volume}
  {62}},\ \bibinfo {pages} {1022--1033} (\bibinfo {year} {2021})}\BibitemShut
  {NoStop}%
\bibitem [{\citenamefont {Meisel}\ \emph {et~al.}(2012)\citenamefont {Meisel},
  \citenamefont {Storch}, \citenamefont {Hallmeyer-Elgner}, \citenamefont
  {Bullmore},\ and\ \citenamefont {Gross}}]{Meisel2012}%
  \BibitemOpen
  \bibfield  {author} {\bibinfo {author} {\bibfnamefont {C.}~\bibnamefont
  {Meisel}}, \bibinfo {author} {\bibfnamefont {A.}~\bibnamefont {Storch}},
  \bibinfo {author} {\bibfnamefont {S.}~\bibnamefont {Hallmeyer-Elgner}},
  \bibinfo {author} {\bibfnamefont {E.}~\bibnamefont {Bullmore}}, \ and\
  \bibinfo {author} {\bibfnamefont {T.}~\bibnamefont {Gross}},\ }\bibfield
  {title} {\enquote {\bibinfo {title} {Failure of adaptive self-organized
  criticality during epileptic seizure attacks},}\ }\href@noop {} {\bibfield
  {journal} {\bibinfo  {journal} {PLoS Comput. Biol.}\ }\textbf {\bibinfo
  {volume} {8}},\ \bibinfo {pages} {e1002312} (\bibinfo {year}
  {2012})}\BibitemShut {NoStop}%
\bibitem [{\citenamefont {Meisel}(2020)}]{Meisel2020EpiSub}%
  \BibitemOpen
  \bibfield  {author} {\bibinfo {author} {\bibfnamefont {C.}~\bibnamefont
  {Meisel}},\ }\bibfield  {title} {\enquote {\bibinfo {title} {Antiepileptic
  drugs induce subcritical dynamics in human cortical networks},}\ }\href
  {\doibase 10.1073/pnas.1911461117} {\bibfield  {journal} {\bibinfo  {journal}
  {Proc. Nat. Acad. Sci. USA}\ }\textbf {\bibinfo {volume} {117}},\ \bibinfo
  {pages} {11118--11125} (\bibinfo {year} {2020})}\BibitemShut {NoStop}%
\bibitem [{\citenamefont {Maturana}\ \emph {et~al.}(2020)\citenamefont
  {Maturana}, \citenamefont {Meisel}, \citenamefont {Dell}, \citenamefont
  {Karoly}, \citenamefont {D’Souza}, \citenamefont {Grayden}, \citenamefont
  {Burkitt}, \citenamefont {Jiruska}, \citenamefont {Kudlacek}, \citenamefont
  {Hlinka} \emph {et~al.}}]{Maturana2020}%
  \BibitemOpen
  \bibfield  {author} {\bibinfo {author} {\bibfnamefont {M.~I.}\ \bibnamefont
  {Maturana}}, \bibinfo {author} {\bibfnamefont {C.}~\bibnamefont {Meisel}},
  \bibinfo {author} {\bibfnamefont {K.}~\bibnamefont {Dell}}, \bibinfo {author}
  {\bibfnamefont {P.~J.}\ \bibnamefont {Karoly}}, \bibinfo {author}
  {\bibfnamefont {W.}~\bibnamefont {D’Souza}}, \bibinfo {author}
  {\bibfnamefont {D.~B.}\ \bibnamefont {Grayden}}, \bibinfo {author}
  {\bibfnamefont {A.~N.}\ \bibnamefont {Burkitt}}, \bibinfo {author}
  {\bibfnamefont {P.}~\bibnamefont {Jiruska}}, \bibinfo {author} {\bibfnamefont
  {J.}~\bibnamefont {Kudlacek}}, \bibinfo {author} {\bibfnamefont
  {J.}~\bibnamefont {Hlinka}},  \emph {et~al.},\ }\bibfield  {title} {\enquote
  {\bibinfo {title} {Critical slowing down as a biomarker for seizure
  susceptibility},}\ }\href@noop {} {\bibfield  {journal} {\bibinfo  {journal}
  {Nat Commun.}\ }\textbf {\bibinfo {volume} {11}},\ \bibinfo {pages} {2172}
  (\bibinfo {year} {2020})}\BibitemShut {NoStop}%
\bibitem [{\citenamefont {\v{Z}iburkus}, \citenamefont {Cressman},\ and\
  \citenamefont {Schiff}(2013)}]{Ziburkus2013EIseizure}%
  \BibitemOpen
  \bibfield  {author} {\bibinfo {author} {\bibfnamefont {J.}~\bibnamefont
  {\v{Z}iburkus}}, \bibinfo {author} {\bibfnamefont {J.~R.}\ \bibnamefont
  {Cressman}}, \ and\ \bibinfo {author} {\bibfnamefont {S.~J.}\ \bibnamefont
  {Schiff}},\ }\bibfield  {title} {\enquote {\bibinfo {title} {Seizures as
  imbalanced up states: excitatory and inhibitory conductances during
  seizure-like events},}\ }\href {\doibase 10.1152/jn.00232.2012} {\bibfield
  {journal} {\bibinfo  {journal} {J. Neurophysiol.}\ }\textbf {\bibinfo
  {volume} {109}},\ \bibinfo {pages} {1296--1306} (\bibinfo {year}
  {2013})}\BibitemShut {NoStop}%
\bibitem [{\citenamefont {Chen}\ \emph {et~al.}(2001)\citenamefont {Chen},
  \citenamefont {Aradi}, \citenamefont {Thon}, \citenamefont {Eghbal{-A}hmadi},
  \citenamefont {Baram},\ and\ \citenamefont {Soltesz}}]{Chen2001seizure}%
  \BibitemOpen
  \bibfield  {author} {\bibinfo {author} {\bibfnamefont {K.}~\bibnamefont
  {Chen}}, \bibinfo {author} {\bibfnamefont {I.}~\bibnamefont {Aradi}},
  \bibinfo {author} {\bibfnamefont {N.}~\bibnamefont {Thon}}, \bibinfo {author}
  {\bibfnamefont {M.}~\bibnamefont {Eghbal{-A}hmadi}}, \bibinfo {author}
  {\bibfnamefont {T.~Z.}\ \bibnamefont {Baram}}, \ and\ \bibinfo {author}
  {\bibfnamefont {I.}~\bibnamefont {Soltesz}},\ }\bibfield  {title} {\enquote
  {\bibinfo {title} {Persistently modified h-channels after complex febrile
  seizures convert the seizure-induced enhancement of inhibition to
  hyperexcitability},}\ }\href {\doibase 10.1038/85480} {\bibfield  {journal}
  {\bibinfo  {journal} {Nat Med}\ }\textbf {\bibinfo {volume} {7}},\ \bibinfo
  {pages} {331--337} (\bibinfo {year} {2001})}\BibitemShut {NoStop}%
\bibitem [{\citenamefont {Naylor}(2010)}]{Naylor2010seizure}%
  \BibitemOpen
  \bibfield  {author} {\bibinfo {author} {\bibfnamefont {D.~E.}\ \bibnamefont
  {Naylor}},\ }\bibfield  {title} {\enquote {\bibinfo {title} {Glutamate and
  gaba in the balance: Convergent pathways sustain seizures during status
  epilepticus},}\ }\href {\doibase 10.1111/j.1528-1167.2010.02622.x} {\bibfield
   {journal} {\bibinfo  {journal} {Epilepsia}\ }\textbf {\bibinfo {volume}
  {51}},\ \bibinfo {pages} {106--109} (\bibinfo {year} {2010})}\BibitemShut
  {NoStop}%
\bibitem [{\citenamefont {Kasteleijn-Nolst~Trenit{\'e}}\ \emph
  {et~al.}(2011)\citenamefont {Kasteleijn-Nolst~Trenit{\'e}}, \citenamefont
  {Rubboli}, \citenamefont {Hirsch}, \citenamefont {Martins~da Silva},
  \citenamefont {Seri}, \citenamefont {Wilkins}, \citenamefont {Parra},
  \citenamefont {Covanis}, \citenamefont {Elia}, \citenamefont {Capovilla},
  \citenamefont {Stephani},\ and\ \citenamefont {Harding}}]{Trenite2011}%
  \BibitemOpen
  \bibfield  {author} {\bibinfo {author} {\bibfnamefont {D.}~\bibnamefont
  {Kasteleijn-Nolst~Trenit{\'e}}}, \bibinfo {author} {\bibfnamefont
  {G.}~\bibnamefont {Rubboli}}, \bibinfo {author} {\bibfnamefont
  {E.}~\bibnamefont {Hirsch}}, \bibinfo {author} {\bibfnamefont
  {A.}~\bibnamefont {Martins~da Silva}}, \bibinfo {author} {\bibfnamefont
  {S.}~\bibnamefont {Seri}}, \bibinfo {author} {\bibfnamefont {A.}~\bibnamefont
  {Wilkins}}, \bibinfo {author} {\bibfnamefont {J.}~\bibnamefont {Parra}},
  \bibinfo {author} {\bibfnamefont {A.}~\bibnamefont {Covanis}}, \bibinfo
  {author} {\bibfnamefont {M.}~\bibnamefont {Elia}}, \bibinfo {author}
  {\bibfnamefont {G.}~\bibnamefont {Capovilla}}, \bibinfo {author}
  {\bibfnamefont {U.}~\bibnamefont {Stephani}}, \ and\ \bibinfo {author}
  {\bibfnamefont {G.}~\bibnamefont {Harding}},\ }\bibfield  {title} {\enquote
  {\bibinfo {title} {Methodology of photic stimulation revisited: updated
  european algorithm for visual stimulation in the {EEG} laboratory},}\ }\href
  {\doibase 10.1111/j.1528-1167.2011.03319.x} {\bibfield  {journal} {\bibinfo
  {journal} {Epilepsia}\ }\textbf {\bibinfo {volume} {53}},\ \bibinfo {pages}
  {16--24} (\bibinfo {year} {2011})}\BibitemShut {NoStop}%
\bibitem [{\citenamefont {Hermes}, \citenamefont
  {Kasteleijn-Nolst~Trenit{\'e}},\ and\ \citenamefont
  {Winawer}(2017)}]{Hermes2017}%
  \BibitemOpen
  \bibfield  {author} {\bibinfo {author} {\bibfnamefont {D.}~\bibnamefont
  {Hermes}}, \bibinfo {author} {\bibfnamefont {D.~G.~A.}\ \bibnamefont
  {Kasteleijn-Nolst~Trenit{\'e}}}, \ and\ \bibinfo {author} {\bibfnamefont
  {J.}~\bibnamefont {Winawer}},\ }\bibfield  {title} {\enquote {\bibinfo
  {title} {Gamma oscillations and photosensitive epilepsy},}\ }\href {\doibase
  10.1016/j.cub.2017.03.076} {\bibfield  {journal} {\bibinfo  {journal} {Curr
  Biol}\ }\textbf {\bibinfo {volume} {27}},\ \bibinfo {pages} {R336--R338}
  (\bibinfo {year} {2017})}\BibitemShut {NoStop}%
\bibitem [{\citenamefont {Honey}\ and\ \citenamefont
  {Valiante}(2017)}]{Honey2017}%
  \BibitemOpen
  \bibfield  {author} {\bibinfo {author} {\bibfnamefont {C.~J.}\ \bibnamefont
  {Honey}}\ and\ \bibinfo {author} {\bibfnamefont {T.}~\bibnamefont
  {Valiante}},\ }\bibfield  {title} {\enquote {\bibinfo {title} {Neuroscience:
  When a single image can cause a seizure},}\ }\href {\doibase
  10.1016/j.cub.2017.03.067} {\bibfield  {journal} {\bibinfo  {journal} {Curr
  Biol}\ }\textbf {\bibinfo {volume} {27}},\ \bibinfo {pages} {R394--R397}
  (\bibinfo {year} {2017})}\BibitemShut {NoStop}%
\bibitem [{\citenamefont {Sayari}\ \emph {et~al.}(2022)\citenamefont {Sayari},
  \citenamefont {Batista}, \citenamefont {Gabrick}, \citenamefont {Iarosz},
  \citenamefont {Hansen}, \citenamefont {Szezech~Jr},\ and\ \citenamefont
  {Borges}}]{sayari2022}%
  \BibitemOpen
  \bibfield  {author} {\bibinfo {author} {\bibfnamefont {E.}~\bibnamefont
  {Sayari}}, \bibinfo {author} {\bibfnamefont {A.~M.}\ \bibnamefont {Batista}},
  \bibinfo {author} {\bibfnamefont {E.~C.}\ \bibnamefont {Gabrick}}, \bibinfo
  {author} {\bibfnamefont {K.~C.}\ \bibnamefont {Iarosz}}, \bibinfo {author}
  {\bibfnamefont {M.}~\bibnamefont {Hansen}}, \bibinfo {author} {\bibfnamefont
  {J.~D.}\ \bibnamefont {Szezech~Jr}}, \ and\ \bibinfo {author} {\bibfnamefont
  {F.~S.}\ \bibnamefont {Borges}},\ }\bibfield  {title} {\enquote {\bibinfo
  {title} {Dynamics of a perturbed random neuronal network with
  burst-timing-dependent plasticity},}\ }\href@noop {} {\bibfield  {journal}
  {\bibinfo  {journal} {Eur Phys J Spec Top}\ }\textbf {\bibinfo {volume}
  {231}},\ \bibinfo {pages} {4049--4056} (\bibinfo {year} {2022})}\BibitemShut
  {NoStop}%
\bibitem [{\citenamefont {Boaretto}\ \emph {et~al.}(2019)\citenamefont
  {Boaretto}, \citenamefont {Budzinski}, \citenamefont {Prado},\ and\
  \citenamefont {Lopes}}]{boaretto2019chialvo}%
  \BibitemOpen
  \bibfield  {author} {\bibinfo {author} {\bibfnamefont {B.}~\bibnamefont
  {Boaretto}}, \bibinfo {author} {\bibfnamefont {R.}~\bibnamefont {Budzinski}},
  \bibinfo {author} {\bibfnamefont {T.}~\bibnamefont {Prado}}, \ and\ \bibinfo
  {author} {\bibfnamefont {S.}~\bibnamefont {Lopes}},\ }\bibfield  {title}
  {\enquote {\bibinfo {title} {Mechanism for explosive synchronization of
  neural networks},}\ }\href@noop {} {\bibfield  {journal} {\bibinfo  {journal}
  {Phys Rev E.}\ }\textbf {\bibinfo {volume} {100}},\ \bibinfo {pages} {052301}
  (\bibinfo {year} {2019})}\BibitemShut {NoStop}%
\bibitem [{\citenamefont {Clark}\ and\ \citenamefont
  {Abbott}(2024)}]{ClarkAbbott2024RateAdapt}%
  \BibitemOpen
  \bibfield  {author} {\bibinfo {author} {\bibfnamefont {D.~G.}\ \bibnamefont
  {Clark}}\ and\ \bibinfo {author} {\bibfnamefont {L.~F.}\ \bibnamefont
  {Abbott}},\ }\bibfield  {title} {\enquote {\bibinfo {title} {Theory of
  coupled neuronal-synaptic dynamics},}\ }\href {\doibase
  10.1103/PhysRevX.14.021001} {\bibfield  {journal} {\bibinfo  {journal} {Phys.
  Rev. X}\ }\textbf {\bibinfo {volume} {14}},\ \bibinfo {pages} {021001}
  (\bibinfo {year} {2024})}\BibitemShut {NoStop}%
\bibitem [{\citenamefont {Sompolinsky}, \citenamefont {Crisanti},\ and\
  \citenamefont {Sommers}(1988)}]{Sompolinsky1988Rate}%
  \BibitemOpen
  \bibfield  {author} {\bibinfo {author} {\bibfnamefont {H.}~\bibnamefont
  {Sompolinsky}}, \bibinfo {author} {\bibfnamefont {A.}~\bibnamefont
  {Crisanti}}, \ and\ \bibinfo {author} {\bibfnamefont {H.~J.}\ \bibnamefont
  {Sommers}},\ }\bibfield  {title} {\enquote {\bibinfo {title} {Chaos in random
  neural networks},}\ }\href {\doibase 10.1103/PhysRevLett.61.259} {\bibfield
  {journal} {\bibinfo  {journal} {Phys. Rev. Lett.}\ }\textbf {\bibinfo
  {volume} {61}},\ \bibinfo {pages} {259--262} (\bibinfo {year}
  {1988})}\BibitemShut {NoStop}%
\bibitem [{\citenamefont {Rosenblum}\ and\ \citenamefont
  {Pikovsky}(2004)}]{Rosenblum2004Feedback}%
  \BibitemOpen
  \bibfield  {author} {\bibinfo {author} {\bibfnamefont {M.}~\bibnamefont
  {Rosenblum}}\ and\ \bibinfo {author} {\bibfnamefont {A.}~\bibnamefont
  {Pikovsky}},\ }\bibfield  {title} {\enquote {\bibinfo {title} {Delayed
  feedback control of collective synchrony: An approach to suppression of
  pathological brain rhythms},}\ }\href {\doibase 10.1103/PhysRevE.70.041904}
  {\bibfield  {journal} {\bibinfo  {journal} {Phys. Rev. E}\ }\textbf {\bibinfo
  {volume} {70}},\ \bibinfo {pages} {041904} (\bibinfo {year}
  {2004})}\BibitemShut {NoStop}%
\bibitem [{\citenamefont {Lameu}\ \emph {et~al.}(2016)\citenamefont {Lameu},
  \citenamefont {Borges}, \citenamefont {Borges}, \citenamefont {Iarosz},
  \citenamefont {Caldas}, \citenamefont {Batista}, \citenamefont {Viana},\ and\
  \citenamefont {Kurths}}]{lameu2016suppression}%
  \BibitemOpen
  \bibfield  {author} {\bibinfo {author} {\bibfnamefont {E.~L.}\ \bibnamefont
  {Lameu}}, \bibinfo {author} {\bibfnamefont {F.~S.}\ \bibnamefont {Borges}},
  \bibinfo {author} {\bibfnamefont {R.~R.}\ \bibnamefont {Borges}}, \bibinfo
  {author} {\bibfnamefont {K.~C.}\ \bibnamefont {Iarosz}}, \bibinfo {author}
  {\bibfnamefont {I.~L.}\ \bibnamefont {Caldas}}, \bibinfo {author}
  {\bibfnamefont {A.~M.}\ \bibnamefont {Batista}}, \bibinfo {author}
  {\bibfnamefont {R.~L.}\ \bibnamefont {Viana}}, \ and\ \bibinfo {author}
  {\bibfnamefont {J.}~\bibnamefont {Kurths}},\ }\bibfield  {title} {\enquote
  {\bibinfo {title} {Suppression of phase synchronisation in network based on
  cat's brain},}\ }\href@noop {} {\bibfield  {journal} {\bibinfo  {journal}
  {Chaos .}\ }\textbf {\bibinfo {volume} {26}} (\bibinfo {year}
  {2016})}\BibitemShut {NoStop}%
\bibitem [{\citenamefont {Mugnaine}\ \emph {et~al.}(2018)\citenamefont
  {Mugnaine}, \citenamefont {Reis}, \citenamefont {Borges}, \citenamefont
  {Borges}, \citenamefont {Ferrari}, \citenamefont {Iarosz}, \citenamefont
  {Caldas}, \citenamefont {Lameu}, \citenamefont {Viana}, \citenamefont
  {Szezech} \emph {et~al.}}]{mugnaine2018delayed}%
  \BibitemOpen
  \bibfield  {author} {\bibinfo {author} {\bibfnamefont {M.}~\bibnamefont
  {Mugnaine}}, \bibinfo {author} {\bibfnamefont {A.~S.}\ \bibnamefont {Reis}},
  \bibinfo {author} {\bibfnamefont {F.~S.}\ \bibnamefont {Borges}}, \bibinfo
  {author} {\bibfnamefont {R.~R.}\ \bibnamefont {Borges}}, \bibinfo {author}
  {\bibfnamefont {F.~A.}\ \bibnamefont {Ferrari}}, \bibinfo {author}
  {\bibfnamefont {K.~C.}\ \bibnamefont {Iarosz}}, \bibinfo {author}
  {\bibfnamefont {I.~L.}\ \bibnamefont {Caldas}}, \bibinfo {author}
  {\bibfnamefont {E.~L.}\ \bibnamefont {Lameu}}, \bibinfo {author}
  {\bibfnamefont {R.~L.}\ \bibnamefont {Viana}}, \bibinfo {author}
  {\bibfnamefont {J.~D.}\ \bibnamefont {Szezech}},  \emph {et~al.},\ }\bibfield
   {title} {\enquote {\bibinfo {title} {Delayed feedback control of phase
  synchronisation in a neuronal network model},}\ }\href@noop {} {\bibfield
  {journal} {\bibinfo  {journal} {Eur Phys J Spec Top}\ }\textbf {\bibinfo
  {volume} {227}},\ \bibinfo {pages} {1151--1160} (\bibinfo {year}
  {2018})}\BibitemShut {NoStop}%
\bibitem [{\citenamefont {Reis}\ \emph {et~al.}(2021)\citenamefont {Reis},
  \citenamefont {Iarosz}, \citenamefont {Ferrari}, \citenamefont {Caldas},
  \citenamefont {Batista},\ and\ \citenamefont {Viana}}]{reis2021bursting}%
  \BibitemOpen
  \bibfield  {author} {\bibinfo {author} {\bibfnamefont {A.~S.}\ \bibnamefont
  {Reis}}, \bibinfo {author} {\bibfnamefont {K.~C.}\ \bibnamefont {Iarosz}},
  \bibinfo {author} {\bibfnamefont {F.~A.}\ \bibnamefont {Ferrari}}, \bibinfo
  {author} {\bibfnamefont {I.~L.}\ \bibnamefont {Caldas}}, \bibinfo {author}
  {\bibfnamefont {A.~M.}\ \bibnamefont {Batista}}, \ and\ \bibinfo {author}
  {\bibfnamefont {R.~L.}\ \bibnamefont {Viana}},\ }\bibfield  {title} {\enquote
  {\bibinfo {title} {Bursting synchronization in neuronal assemblies of
  scale-free networks},}\ }\href@noop {} {\bibfield  {journal} {\bibinfo
  {journal} {Chaos Solitons Fractals}\ }\textbf {\bibinfo {volume} {142}},\
  \bibinfo {pages} {110395} (\bibinfo {year} {2021})}\BibitemShut {NoStop}%
\end{thebibliography}
\providecommand{\noopsort}[1]{}\providecommand{\singleletter}[1]{#1}%
%


\newpage\clearpage
\onecolumngrid
\appendix
\renewcommand{\thefigure}{S\arabic{figure}}
\renewcommand{\thetable}{S\arabic{table}}
\setcounter{figure}{0}
\setcounter{table}{0}

\section{Supplementary Information}
This document contains extra figures and supporting information about the methods applied in the manuscript:
\begin{itemize}
    \item We display the interspike interval distributions and power spectra of a single randomly selected neuron connected
    to the all-to-all network at and around the synchronization phase transition.
    \item  We also explain in details the fitting of the damping time scale
of the reverberating activity around the critical point.
\end{itemize}

\section{The isolated neuron}

\begin{table}[!ph]
    \centering
    \begin{tabular}{c|ccccccc}
    \thickhline
       \sf{\textbf{Label}} & $K$   &  $T$   & $\delta_i$ &   $u$   & $\epsilon$ &   $H$   & $I_i(t)$\\
    \toprule
       \sf{\textbf{S1}}    & $0.6$ & $0.35$ & $0.006$    & $0.004$ & $-0.98$    & $-0.5 $ &  $0$\\
       \sf{\textbf{S2}}    & $0.6$ & $0.35$ & $0.006$    & $0.004$ & $-0.98$    & $-0.2 $ &  $-0.06$\\
       \sf{\textbf{CS}}    & $0.6$ & $0.35$ & $0.006$    & $0.004$ & $-0.98$    & $-0.95$ &  $0$\\
       \sf{\textbf{B1}}    & $0.6$ & $0.35$ & $0.006$    & $0.004$ & $-0.98$    & $-0.8 $ &  $0$\\
       \sf{\textbf{B2}}    & $0.7$ & $0.5 $ & $0.008$    & $0.004$ & $-0.98$    & $-0.9 $ &  $0$\\
       \sf{\textbf{B3}}    & $0.8$ & $0.6 $ & $0.006$    & $0.004$ & $-0.99$    & $-0.8 $ &  $0$\\
    \thickhline
    \end{tabular}
    \caption{Parameters for the attractors in Fig.~1 of the main text.}
    \label{tab:attractorparam}
\end{table}

\section{Microscopic activity characterization}

\begin{figure*}[!htbp]
    \centering
    \includegraphics[width=\textwidth]{{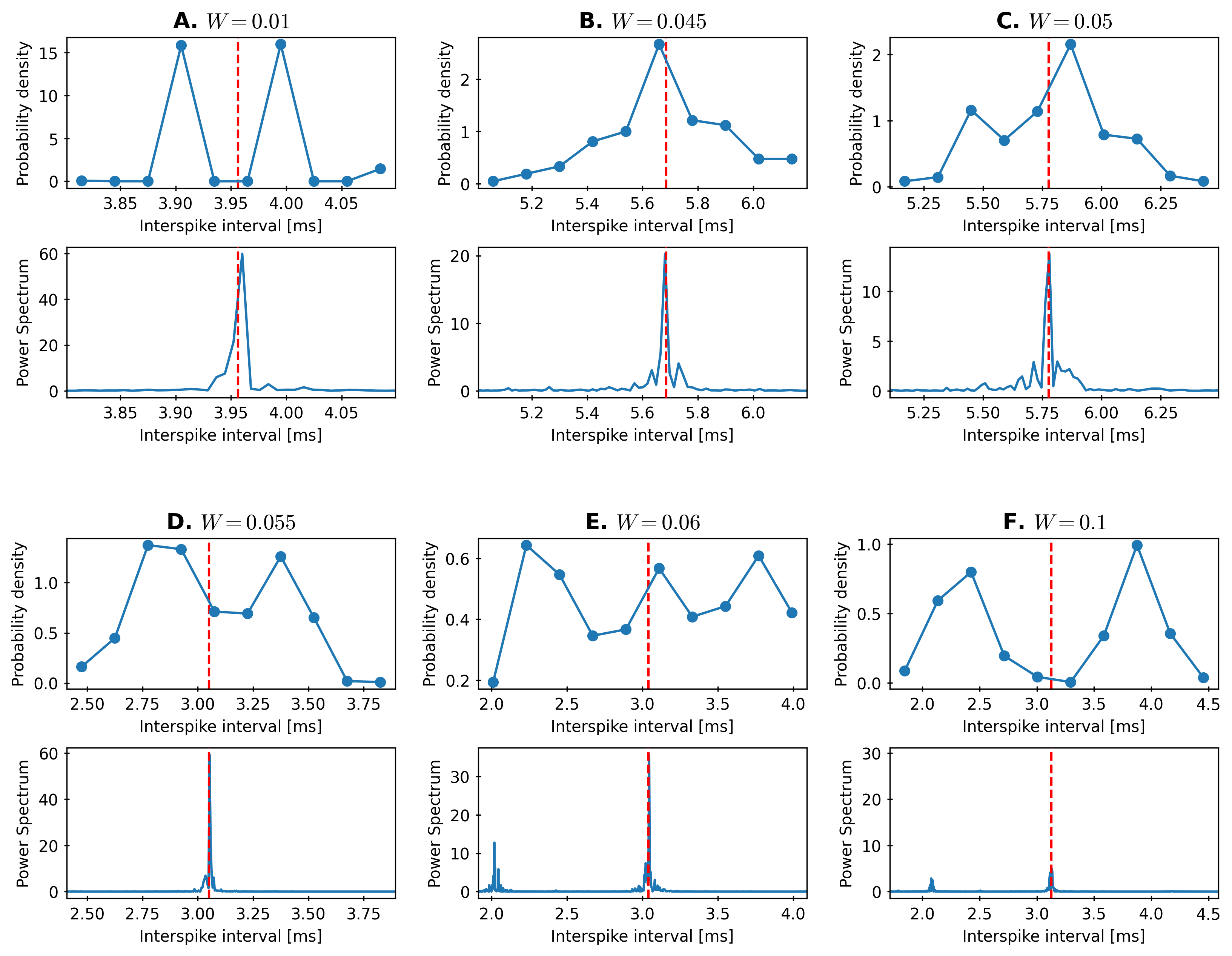}}
    \caption{\label{fig:freqDist}
    \textbf{Interspike interval of a neuron inside the network.}
    A network of $N=1000$ neurons is simulated for each $W$ around the synchronization phase transition,
    and we characterize the activity of a single neuron within it.
    The interspike inverval (ISI) distributions are broad and bimodal in both asynchronous and synchronous phases,
    although around the transition range ($0.045\lesssim W\lesssim0.05$) it becomes unimodal, assuming $W_c\approx0.045$ (see the main text).
    Legend: \red{\textbf{- - -}} $\avg{ISI}$. Notice that $\avg{ISI}$ peaks along the transition range of $W$.
    It goes from $\avg{ISI}\approx4$~ms for $W<W_c$, to $\avg{ISI}\approx5.7$~ms at $W\approx W_c$, and then returns to $\avg{ISI}\approx3$~ ms for $W>W_c$.
    \textbf{A.} $W=0.01$.
    \textbf{B.} $W=0.045$.
    \textbf{C.} $W=0.05$.
    \textbf{D.} $W=0.055$.
    \textbf{E.} $W=0.06$.
    \textbf{F.} $W=0.1$.
    Parameters: $K=0.6$, $T=0.35$, $H=-0.2$, $\delta=0.006$, $\Delta=0.003$, $u=0.004$, $\epsilon=-0.98$.
    }
\end{figure*}

\FloatBarrier

\section{Fitting the damping of the firing rate near the critical point}

The firing rate of the network $\rho[t]$ is simply the average count of spikes per time step,
\begin{equation}
    \rho[t] = \dfrac{1}{N}\sum_{i=1}^N S_i[t]\:,
\end{equation}
where $S_i[t]=1$ if a spike happened at time $t$ ($S_i[t]=0$ otherwise). See the manuscript for the spike detection condition.
We simulate the network until it reaches the stationary state, then continue in the stationary state for 1000 time steps (100 ms) before
injecting the stimulus at $t_0=100$~ms.
The time average $\avg{\rho}$ is taken over the pre-stimulus stationary state, $t<t_0$. 
The difference $\rho-\avg{\rho}$ is the detrended firing rate signal. From the observation of this signal, we assume that it
can be written as the following function of the delay $\Delta t=t-t_0>0$
\[
\left(\rho-\avg{\rho}\right)[\Delta t]=A[\Delta t] C[\Delta t]\ ,
\] where $A[\Delta t]$ is the time-dependent amplitude of the large oscillations $C[\Delta t]$.

\begin{figure*}[!tp]
    \centering
    \includegraphics[width=\textwidth]{{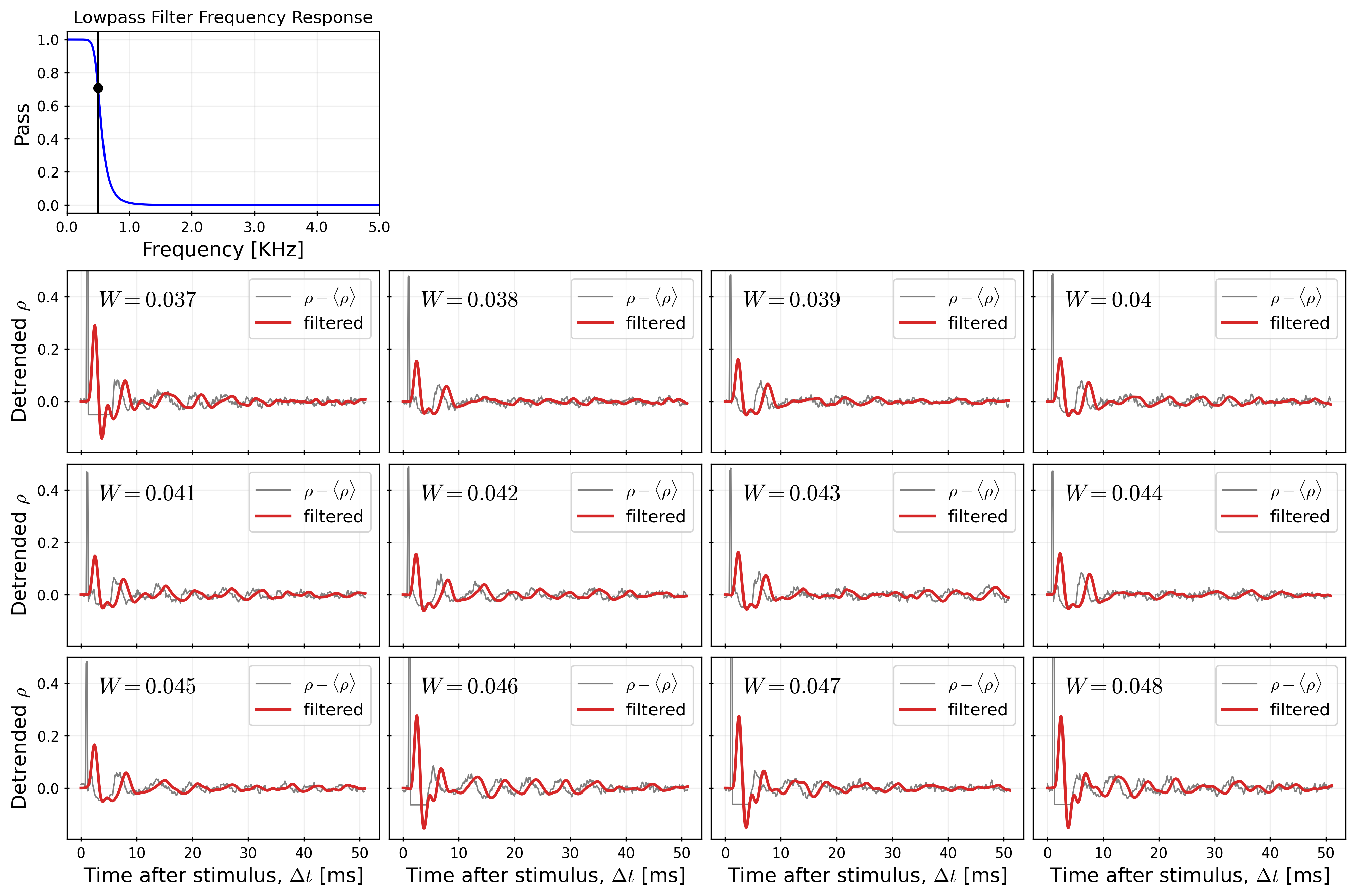}}
    \caption{\label{fig:filter}
    \textbf{Firing rate low-pass filter.}
    \textbf{Top.} Filter response function showing a smooth cutoff of frequencies at $500$~KHz. This filter is applied to the post-stimulus
detrended firing rate oscillations of the network to produce a clean decay of the larger (and slower) oscillations.
    \textbf{Other panels.} Raw and filtered detrended firing rate $\rho-\avg{\rho}$ as a function of the delay after stimulus $\Delta t$.
    The filtered signal is slightly delayed, but its amplitude and damping match the noisy oscillations.
    The initial raw spike in the raw rate is the network response during the stimulus.
    $W$ as given in the panels.
    Parameters: $N=1000$, $K=0.6$, $T=0.35$, $H=-0.2$, $\delta=0.006$, $\Delta=0.003$, $u=0.004$, $\epsilon=-0.98$.
    }
\end{figure*}

\begin{figure*}[!tp]
    \centering
    \includegraphics[width=\textwidth]{{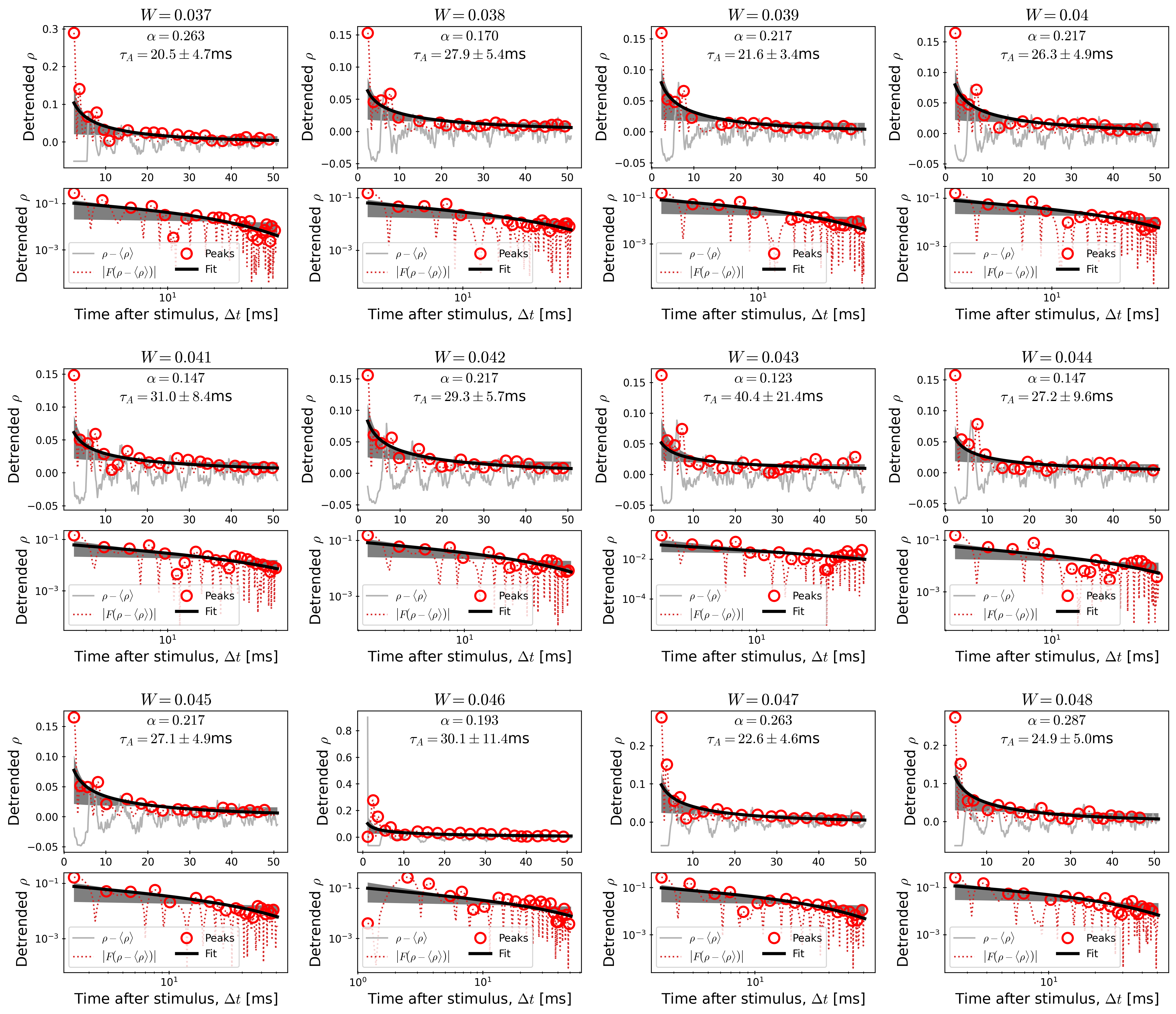}}
    \caption{\label{fig:ampfit}
    \textbf{Fitting the amplitude of the large and slow oscillations.}
The panels are presented in doubles (top and bottom of each of the three rows): for each $W$, the same data are shown on a linear scale (top of the row) and
on a log-log scale (bottom of the row). We extracted the peaks of the absolute value of the filtered detrended firing rates, $|F(\rho-\avg{\rho})|$, that had prominence greater than 0.002 (firing rate units) -- shown in circles. We fitted Eq.~\eqref{eq:ampdecayfunc} for multiple $\alpha$ (remained fixed during fitting) to obtain the time scale $\tau_A$ of the damping. We show the $\alpha$ that minimized the error of $\tau_A$ (bold black line). Gray shading: Fit error range.
    $W$ as given in the panels.
    Parameters: $N=1000$, $K=0.6$, $T=0.35$, $H=-0.2$, $\delta=0.006$, $\Delta=0.003$, $u=0.004$, $\epsilon=-0.98$.
    }
\end{figure*}

Now we describe the procedure to fit the amplitude $A$. This is applied independently to the detrended firing rate
data of the simulated network for each $W$. First, we need to eliminate fast stochastic fluctuations from the detrended firing rate.
Thus, we define a standard low-pass filter $F(u)$ with a soft threshold at $500$ KHz (Fig.~\ref{fig:filter}),
and $u=\left(\rho-\avg{\rho}\right)[\Delta t]$ is the function of time to be filtered.
We take the absolute value of the filtered signal $|F(\rho-\avg{\rho})|$ in order to convert the wave valleys into peaks.
We select peaks with prominence greater than 0.002 (in firing rate units, Fig.\ref{fig:ampfit}) and use them to fit the amplitude damping function:
\begin{equation}
\label{eq:ampdecayfunc}
    A[\Delta t]=B\Delta t^{-\alpha}\exp\!\left(-\Delta t/\tau_A\right)\:.
\end{equation}
$B$ is just a fitting constant, $\tau_A$ is the characteristic time of the damping during the post-stimulus delay $\Delta t$.
$\alpha$ is a fixed parameter.

We sweep $\alpha$ over the range $[0.1;0.8]$, and then for each $\alpha$ we perform a non-linear least squares fit of Eq~\eqref{eq:ampdecayfunc} to the
amplitude peaks data and obtain $B(\alpha)$ and $\tau_A(\alpha)$ as functions of $\alpha$. The statistical error $E_{\tau_A}$ of the fitted $\tau_A$ is also a function of $\alpha$.
We minimize $E_{\tau_A}$ to obtain the best fit:
\begin{equation}
    \alpha_{\rm best} = \underset{\alpha}{\rm argmin}\!\left[E_{\tau_A}(\alpha)\right]\:,
\end{equation}
yielding $\tau_{A,{\rm best}}=\tau_A(\alpha_{\rm best})$. $\tau_{A,{\rm best}}$ is the value shown in the panels of Fig.~\ref{fig:ampfit}.

Since this is done independently for each $W$, the best fit values $\tau_{A,{\rm best}}$ and $\alpha_{\rm best}$ are functions of $W$ (shown in Fig.~5C of the main text
without the ``best'' subscript for clarity).
To confirm the optimization of $\tau_A$, we can replace $\alpha$ in Eq.~\eqref{eq:ampdecayfunc} by its average best value $\alpha=\bar{\alpha}_{\rm best}=0.20\pm0.05$,
and again run the fit of $\tau_A$ and $B$. This results in the same behavior for $\tau_A$ as a function of $W$; \textit{i.e.},
$\tau_A$ is optimized at $W=0.043$.

\end{document}